\documentclass[namedreferences]{solarphysics}

\usepackage[hyperref,optionalrh,showbiblabels]{spr-sola-addons} 
\usepackage{graphicx}        
\usepackage{color}           
\usepackage{breakurl}        

\newcommand{\aap}{    {\it Astron. Astrophys.}}

\newcommand{\apj}{    {\it Astrophys. J.}}

\newcommand{\jgr}{    {\it J. Geophys. Res.}}

\newcommand{\solphys}{{\it Solar Phys.}}

\chardef\us=`\_

\begin{document}

\begin{article}
\begin{opening}

\title{Magnetic reconnections in the presence of three-dimensional magnetic nulls and quasi-separatrix layers}

\author[addressref={aff1},corref,email={sainisanjay35@gmail.com}]{\inits{S.}\fnm{Sanjay}~\lnm{Kumar}}
\author[addressref={aff2,aff3}]{\inits{S. S.}\fnm{Sushree S}~\lnm{Nayak}}
\author[addressref={aff4}]{\inits{A.}\fnm{Avijeet}~\lnm{Prasad}}
\author[addressref=aff2]{\inits{R.}\fnm{Ramit}~\lnm{Bhattacharyya}}

\address[id=aff1]{Department of Physics, Patna University, Patna-80005, India}
\address[id=aff2]{Udaipur Solar Observatory, Physical Research Laboratory, Dewali, Bari Road, Udaipur-313001, India}
\address[id=aff3]{Department of Physics, Indian Institute of Technology, Palaj, Gandhinagar, 382355, India}
\address[id=aff4]{Center for Space Plasma and Aeronomic Research, The University of Alabama in Huntsville, Huntsville, AL 35899, USA.}

\runningauthor{Kumar et al.}
\runningtitle{Reconnections at nulls and QSLs}

\begin{abstract}
 Three-dimensional (3D) magnetohydrodynamic simulations are carried out to explore magnetic reconnections in the presence of 3D magnetic nulls and quasi-separatrix layers (QSLs). The initial magnetic fields are created
by superposing uniform vertical magnetic fields of two different magnitudes on a linear force-free field. The interior of the numerical box contains two 3D nulls with separatrix domes separated by a quasi-separator (or hyperbolic flux tube) with QSLs. In the first simulation, the uniform vertical field is so large that the nulls are located at low heights and the domes are separate. Initially unbalanced Lorentz forces drive rotational flows that form strong electric currents and strong torsional fan reconnection at the 3D nulls and weak QSL reconnection at the hyperbolic flux tube. Flipping or slipping of field lines is observed in both cases.
In the second simulation, with a weaker vertical field and larger domes, the separatrix surfaces meet at the central quasi-separator and their rotation drives stronger QSL reconnection than before.

\end{abstract}
\end{opening}

\section{Introduction}
     \label{S-Introduction} 
The Sun exhibits transient release of energy through a myriad of phenomena like flares, coronal mass ejections (CMEs), coronal jets, etc. 
The magnitudes of the energy release for solar flares and coronal jets vary widely, with the jets being roughly $10^5$ times less energetic than typical large solar flares, which are known to release energy around $10^{32-33}$ ergs  \citep{aschwanden, priest-new}. The onset of these transients is attributed to the release of magnetic energy stored in the coronal magnetic field \citep{2011LRSP....8....6S, priest-new}. However, the cause of such a catastrophic release of the magnetic energy is still not fully understood. In this direction, the widely accepted physical process is magnetic reconnection (MR) --- a diffusive process in which magnetic energy stored in the plasma gets converted into kinetic energy, heat, and fast particle energy accompanied by a change in magnetic topology  \citep{2011LRSP....8....6S, priest-new, sanjay-2014}. 

Toward identifying the favorable sites for reconnection, 
important are the three-dimensional (3D) magnetic nulls {\citep{2005PhPl...12e2307P, hachami-2010, sanjay-2014, sanjay-cont}} which are the locations where the magnetic field ${\bf{B}}=0$. Coronal magnetic field obtained by extrapolating photospheric magnetic field using various models shows the existence of 3D nulls in abundance in the corona \citep{2009SoPh..254...51L, pontin-2016, prasad+2018apj, 2019ApJ...875...10N}. The characteristics of a coronal 3D null are its  spine and dome-shaped fan structures which owe their origin to the
avoidance of the ${\bf{B}}=0$ location by the magnetic field lines (MFLs).  Because of the fan-spine structure, MRs naturally commence at the 3D nulls and, are commonly known as the null-point magnetic reconnections \citep{pontin-2013, pontin-2016, prasad+2018apj, 2020ApJ...892...44N, prasad-2020}. Recent data-constrained magnetohydrodynamic (MHD) simulations attribute the generation of circular flare ribbons and blowout coronal jets \citep{2009ApJ...700..559M, 2012ApJ...760..101W, 2019ApJ...875...10N, liu-2020, prasad-2020} along with confined flares to the null-point MRs \citep{2007ApJ...662.1293U, prasad+2018apj}. Null point reconnection can occur in three ways at a separatrix dome, namely, spine-fan reconnection when the null collapses to form a current layer, and torsional spine or fan reconnection with rotational motions \citep{2009PhPl...16l2101P, prasad+2018apj, 2019ApJ...875...10N}.

 \citet{schindler1988} suggested that the magnetic nulls are not the only locations in the solar corona where the magnetic reconnection can take place by documenting the flaring events having similar characteristics in the presence and the absence of the magnetic nulls. Two other coronal locations have been proposed for reconnection, namely, separators and quasi-separators (or hyperbolic flux tubes, \citet{2002JGRA..107.1164T}). Separators \citep{lau1990, priest-titov1996, longcope-cowley1996, parnell-etal2010}
 represent the intersection of two separatrix surfaces at which the field line mapping is singular, and have been shown to be present in some flares \citep{longcope-etal2007, titov-etal2012}. Quasi-separators represent the intersection of two quasi-separatrix layers (QSLs).
They were first proposed by \citet{1995JGR...10023443P} and studied further by \citet{1996A&A...308..643D, 2006AdSpR..37.1269D, janvier-2015, pontin-2016, liu-2016}. QSLs are mathematically defined by calculating the field line mapping of the magnetic field suggested by \citet{1995JGR...10023443P} and with an improved measure (the Q factor) discovered by  \citet{2002JGRA..107.1164T}. Using data based MHD simulations, the correspondence between flare ribbons and QSL locations is identified  which shows that the QSLs are also preferential sites for MRs \citep{demoulin-1997, 2009ApJ...700..559M,  janvier-2015, 2017JPlPh..83a5301J, prasad+2018apj}. 
 3D MHD simulations having separators or QSLs confirm that MFLs near them can slip through the plasma and undergo repetitive MRs by exchanging their connectivity with neighboring field lines {\citep{2005PhPl...12e2307P, 2006SoPh..238..347A, aulanier-2007, 2006A&A...451.1101D,  2006A&A...459..627D, prasad+2018apj}}.  

Additional to the pre-existing preferential locations such as nulls, separators and QSLs, MHD simulations with idealized scenario of infinite electrical conductivity show that the potential sites for reconnection can naturally be generated because of the inherent dynamics \citep{sanjay-2014, sanjay-2015}. The magnetofluid evolution being congruent with the Parker's magnetostatic theorem \citep{1972ApJ...174..499P, parker-book, 2012PPCF...54l4028P}, attributes the generation of such sites to a development of favorable magnetic stresses which naturally bring  non-parallel magnetic field lines in close proximity. As a result, layers of intense volume current density ${\bf{J}}$ originate --- known in the literature as current sheets (CSs). In the presence of slight but non-zero magnetic diffusivity, which is the case for the coronal plasma, the CSs get dissipated by magnetic reconnections.   
 
 In the above background, the current paper presents MHD simulations aiming to numerically explore magnetic reconnections, initiated by the presence of 3D nulls and QSLs; and to examine the role of such MRs in shaping-up the dynamics.
 To have a better control over the initial magnetic topology, we utilize analytically constructed initial magnetic fields which have 3D magnetic nulls --- morphologically similar to the ones observed in the solar corona --- and QSLs. Additionally, the  magnetic fields are envisaged to support the Lorentz force to naturally initiate dynamics without any prescribed boundary flow. The simulated MHD evolutions document the MRs at the 3D nulls and QSLs. An important finding of the presented simulations  is in the realization that just an existence of a null, separator or QSL does not guarantee the onset of energetically efficient reconnections, but the nature of the flows in the neighbourhood of such geometric structures
 is equally important --- as previously theorized by \citet{priest-forbes1989}. Further, the simulations identify the autonomous development of current sheets and consequent MRs.

The paper is organized as follows. The initial magnetic field is described in Section 2. The governing MHD equations and numerical model are discussed in Sections 3.  Results of the simulation are presented in Section 4. Section 5 summarizes these results and discusses the key findings.

\section{Initial Magnetic Field} 
      \label{ivps}      

To achieve a complex magnetic topology with 3D nulls and QSLs, the initial magnetic field is constructed by modifying 
the field in {\citet{sanjay-cyl}} for its simple construction based on the superposition of a constant vertical  field  over a linear force-free field (LFFF), defined 
in a Cartesian domain. Particularly, to make MFLs more relevant to the solar corona, we consider the magnetic field to exponentially decay along the z-direction
 in the positive half-space $\Gamma$ defined by $( z \ge 0 )$, instead of the periodic magnetic field assumed in the original configuration. Consequently, the $z=0$ plane is treated as the photosphere. Moreover, the constant vertical magnetic field with straight MFLs is not likely to alter the topology of the LFFF appreciably and, hence, the superposed field $\bf{B}$ is expected to be topologically similar to the unperturbed LFFF.  Relevantly, the solar corona is thought to be in the state of force-free equilibrium under the approximation of a thermodynamic pressure that is negligibly small compared to the magnetic pressure  {\citep{priest-new}}.

As proposed, the initial magnetic field is derived by superposing a 3D LFFF ${\bf{B}_1}$ and a uniform vertical field ${\bf{B}_2}$, where the components of ${\bf{B}_1}$ are,

\begin{eqnarray}
& & {B_{1x}}=  \sin\left(x-y\right) \exp\left( -z\right),
 \label{bx1}
 \\
& & {B_{1y}}= -\sin\left(x+y\right) \exp\left( -z\right),
 \label{by1}
 \\
& & {B_{1z}}=2 \sin\left( x \right) \sin\left( y\right) \exp\left(-z \right).
\label{bz1}
\end{eqnarray}

\noindent The magnetic circulation per unit flux of ${\bf{B}_1}$ has a value of unity
and measures the twist of the corresponding MFLs \citep{parker-book, sanjay-2014}.  The superposed field ${\bf{B}}$ is

\begin{eqnarray} 
\label{eq1}
{\bf{B}}={\bf{B}_1}+c_0{\bf{B}_2}, 
\end{eqnarray}

\noindent where the superposition coefficient $c_0$ relates the amplitudes of the two superposing fields and determines the deviation of ${\bf{B}}$  from the force-free equilibrium \citep{sanjay-cyl}. Explicitly, 

\begin{eqnarray}
\label{super}
& & {B_{x}}=  \sin\left(x-y\right) \exp\left( -z\right),
 \label{bx}
 \\
& & {B_{y}}= -\sin\left(x+y\right) \exp\left( -z\right),
 \label{by}
 \\
& & {B_{z}}=2 \sin\left( x \right) \sin\left( y\right) \exp\left(-z \right)+c_0,
\label{bz}
\end{eqnarray}

\noindent in the domain $\Gamma$ which, physically extends  from $0$ to $2\pi$ while having periodic and open boundaries in the lateral  ($x$ and $y$) and the vertical  directions, respectively. As the LFFF $\bf{B}_1$ being exponentially decaying along $z$, all the three components of $\bf{B}$ are also exponentially decaying functions along the vertical. 
 
The Lorentz force is
\begin{eqnarray}
{\bf{J}} \times {\bf{B}} = c_0 ({\bf{B}}_1 \times {\bf{B}}_2)
\label{lorentz}
\end{eqnarray}

\noindent which is non-zero for $c_0\neq 0$ and has the functional form 

\begin{eqnarray}
\label{superlor}
& & {({\bf{J}} \times {\bf{B}})_{x}}= - c_0\sin\left(x+y\right) \exp\left( -z\right),
 \label{lorx}
 \\
& & {({\bf{J}} \times {\bf{B}})_{y}}= -c_0\sin\left(x-y\right) \exp\left( -z\right),
 \label{lory}
 \\
& & {({\bf{J}} \times {\bf{B}})_{z}}= 0.
\label{lorz}
\end{eqnarray}

\noindent Clearly, the initial Lorentz force  acts laterally. For the simulations, we select $c_0=0.1$ and $c_0=0.5$ to obtain two sets of initial magnetic fields with different magnitudes of Lorentz force --- allowing us to assess the role of different dynamical evolution of the MFLs on the MRs.

\begin{figure}[htp]
\centering
\includegraphics[angle=0,scale=.21]{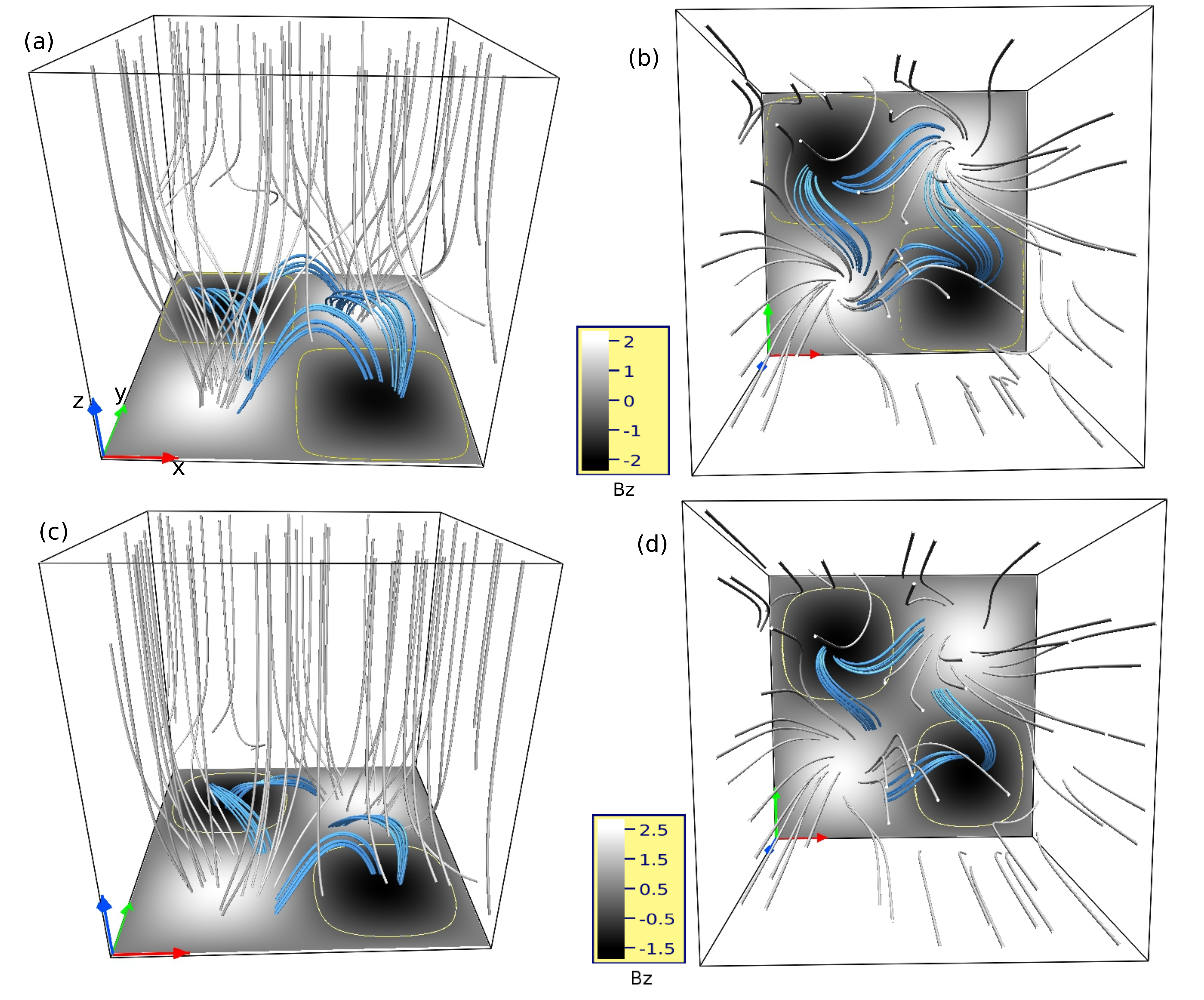}
\caption{Side (panel (a)) and top (panel (b)) views of MFLs of the ${\bf{B}}$ for $c_0=0.1$. While panels (c) and (d) illustrate side and top views of the MFLs for $c_0=0.5$. The MFLs are in the form of twisted closed (marked by navy blue color) as well as open magnetic loops (shown in grey color). All the panels are overplotted with  $B_z$ values on the $z=0$ plane. Yellow line represents the contour corresponding $B_z =0$. } \label{newfigure1}
\end{figure}

To explore the geometrical similarity of the initial MFLs with the coronal MFLs, in Figure \ref{newfigure1} we depict the MFLs of ${\bf{B}}$ for the cases $c_0=0.1$ (panels (a) and (b)) and $c_0=0.5$ (panels (c) and (d)). The figure shows a physical resemblance of the MFLs to the open and the closed coronal loops.

\begin{figure}[htp]
\centering
\includegraphics[angle=0,scale=.21]{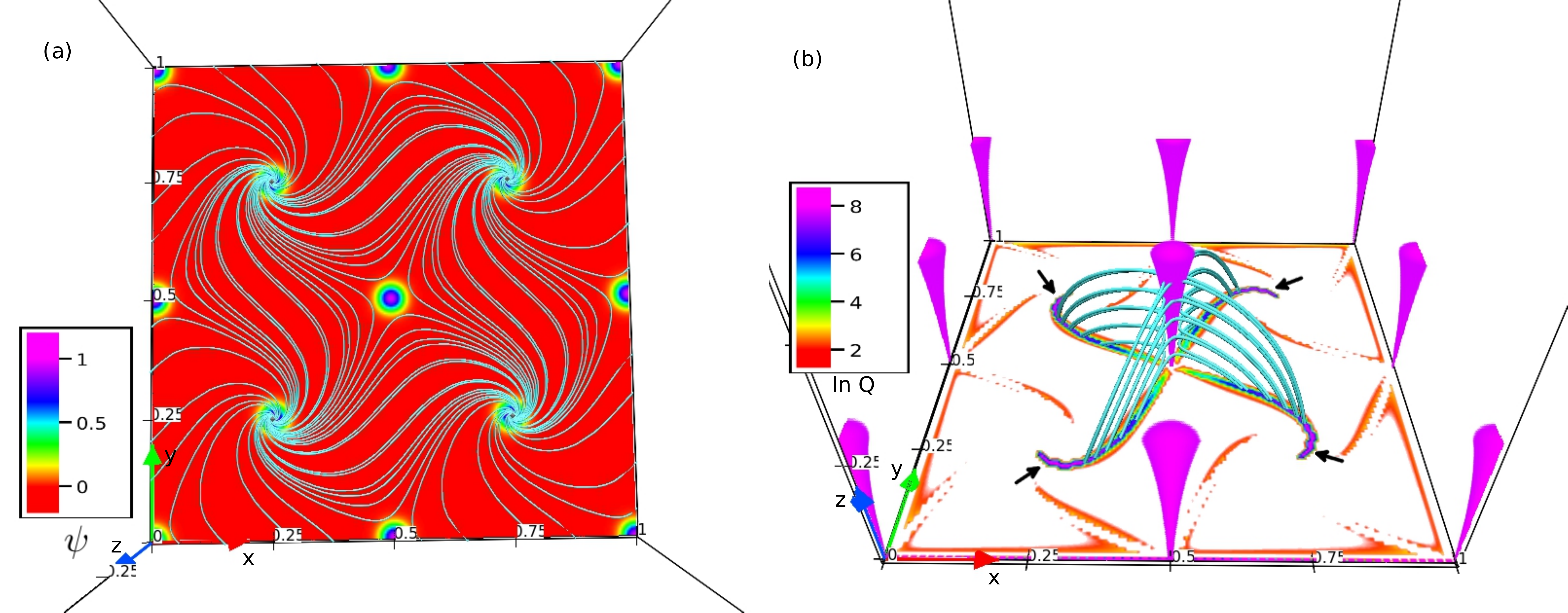}
\caption{ Panel (a) shows the neutral points and the field lines of the transverse field overlaid with the function $\psi$ at $z=0$ plane. Large values of $\psi$ represent the locations of neutral points. Notable is the existence of the nine X-type (one inside the domain and eight at the boundaries) and four spiral-type neutral points. Panel (b) depicts the magnetic nulls (in pink) and MFLs of ${\bf{B}_1}$ overlaid with the corresponding $Q$-map at $z=0$. The nine X-type neutral points of the transverse field also retain in ${\bf{B}_1}$. Noticeable is the larger values of $Q$ near the sites of the spiral neutral points (marked by black arrows) --- suggesting that the spiral neutral points correspond to QSLs in ${\bf{B}_1}$. In the figure, the domain size is marked as 1 instead of $2\pi$ in all the directions.    } \label{newfigure2}
\end{figure}

To carefully examine the magnetic topology of the initial field ${\bf{B}}$, first we plot neutral points in its transverse field (obtained by setting $B_z=0$ in ${\bf{B}}$). Notably, in all relevant illustrations, the neutral point is depicted by using the numerical technique documented in \citet{2020ApJ...892...44N}.  Succinctly, the technique utilizes a Gaussian function  
$\psi=\rm{exp}\left[-\sum_{i=x,y,z}(\textit{B}_i-\textit{B}_0)^2/\textit{d}_0^2\right] $, where $B_0$ and $d_0$ are constants defining a particular isovalue of $B_i$ and the width of the Gaussian respectively. By choosing $B_0 \approx 0$ and a small $d_0$, the function $\psi$ takes significant values only if $B_i\approx 0$ for each $i$. A 3D null is then the point where the three isosurfaces having isovalues $B_i=B_0$ intersect. Figure \ref{newfigure2}(a) shows the neutral points in the transverse field overlaid with corresponding field lines at $z=0$ plane. Notably, the field line geometry near these neutral points suggest that there are four spiral-type nulls {\citep{lau1990}} at 
$(x, y)=(\pi/2, \pi/2)$, $(\pi/2, 3\pi/2)$, $(3\pi/2, \pi/2)$, $(3\pi/2, 3\pi/2)$, and one X-type null {\citep{sanjay-2015}} at $(x, y)=(\pi, \pi)$ inside the computational box.  To further verify, we have checked that the eigenvalues for the X-type null are real ($\sqrt{2}, -\sqrt{2}$) and, for the spiral nulls are complex numbers (for example the eigenvalues of a spiral null at $(\pi/2, \pi/2)$ are ($1+i, 1-i$)). In addition to these five nulls, there are eight X-type nulls at the boundaries of the domain. Relevantly, MRs can occur on separators with spiral-type as well as X-type neutral points in the perpendicular plane {\citep{parnell-etal2011}}.
Next we note that the superposition of  $B_{1z}$ (Equation \ref{bz1}) on the transverse field generates ${\bf{B}_1}$. In Figure \ref{newfigure2}(b), we illustrate the magnetic nulls in ${\bf{B}_1}$ overplotted with its MFLs. Nine X-type neutral lines are evident in ${\bf{B}_1}$ which are co-located with the X-type nulls of the transverse field at $z=0$ plane. However, the four spiral nulls get destroyed in ${\bf{B}_1}$. To relate the location of the spiral nulls to the possible QSLs, we also plot the $Q$-map at the bottom boundary in Figure \ref{newfigure2}(b) by using the code of \citet{liu-2016} available at \url{http://staff.ustc.edu.cn/~rliu/qfactor.html}. The same code is used to plot Q-map throughout the paper.
Notably, the regions with large $Q$-values include both separatrices and QSLs \citep{2002JGRA..107.1164T}.  Important are the large $Q$-values at the locations of the spiral nulls (marked by black arrows in Fig. \ref{newfigure2}(b)), suggesting that some of the spiral nulls (of the transverse field) convert into the QSLs for ${\bf{B}_1}$.

\begin{figure}[htp]
\centering
\includegraphics[angle=0,scale=.21]{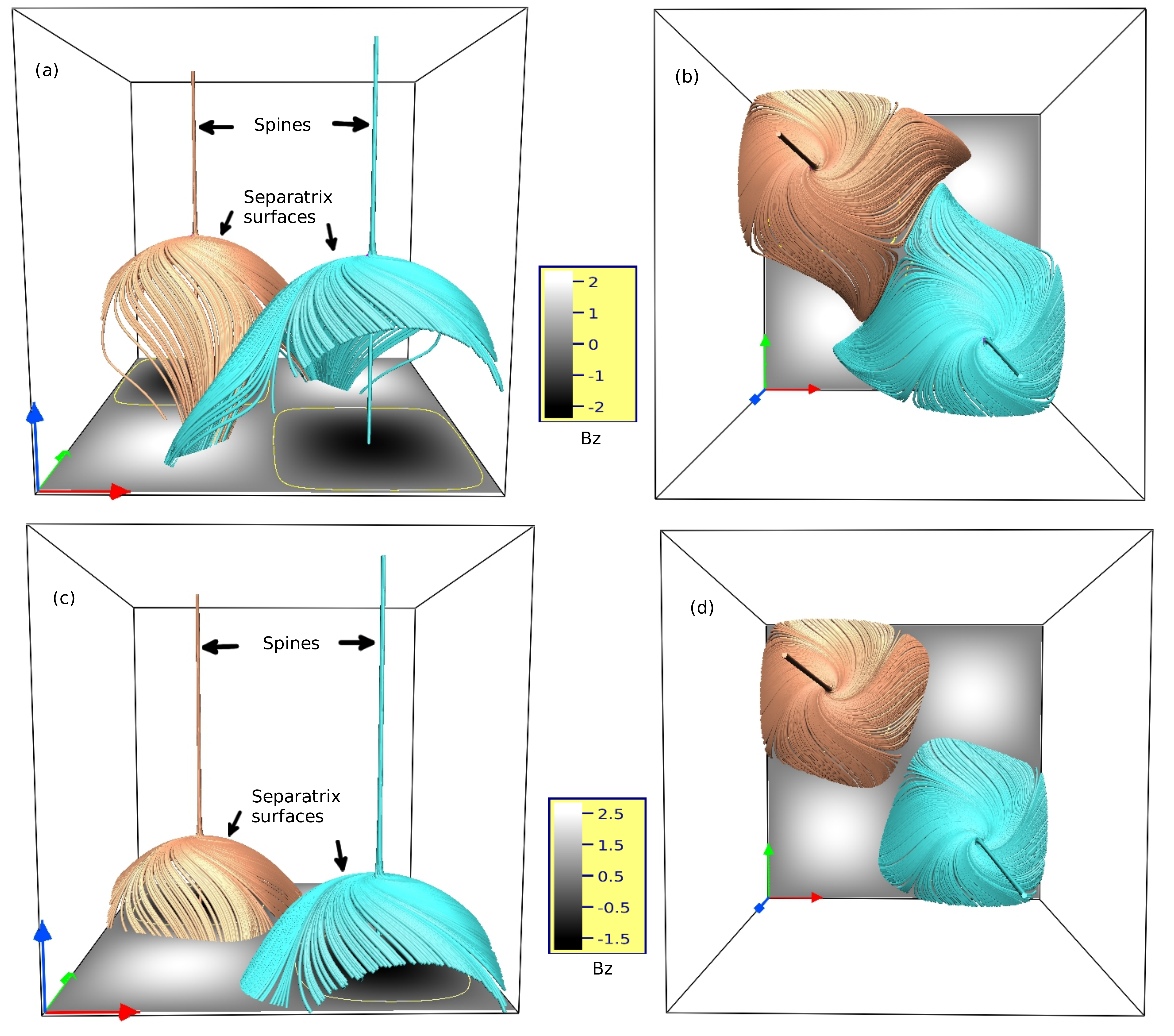}
\caption{ Magnetic (topological) skeleton of the initial field $\bf{B}$ in terms of 3D nulls, separatrix surfaces and spines for $c_0=0.1$ (side view in panel (a) and top view in panel (b)) and $c_0=0.5$ (side view in panel (c) and top view in panel (d)). Notable are straight spine axes and the dome-shaped separatrix or fan surfaces (intersecting bottom boundary) of the nulls. For $c_0=0.1$, the separatrix domes touch at the base of the quasi-separator at $(\pi, \pi, 0)$ (panel (b)). For $c_0=0.5$, the separatrix domes are separate (panel (d)).}  \label{newfigure3}
\end{figure}

\begin{figure}[htp]
\centering
\includegraphics[angle=0,scale=.21]{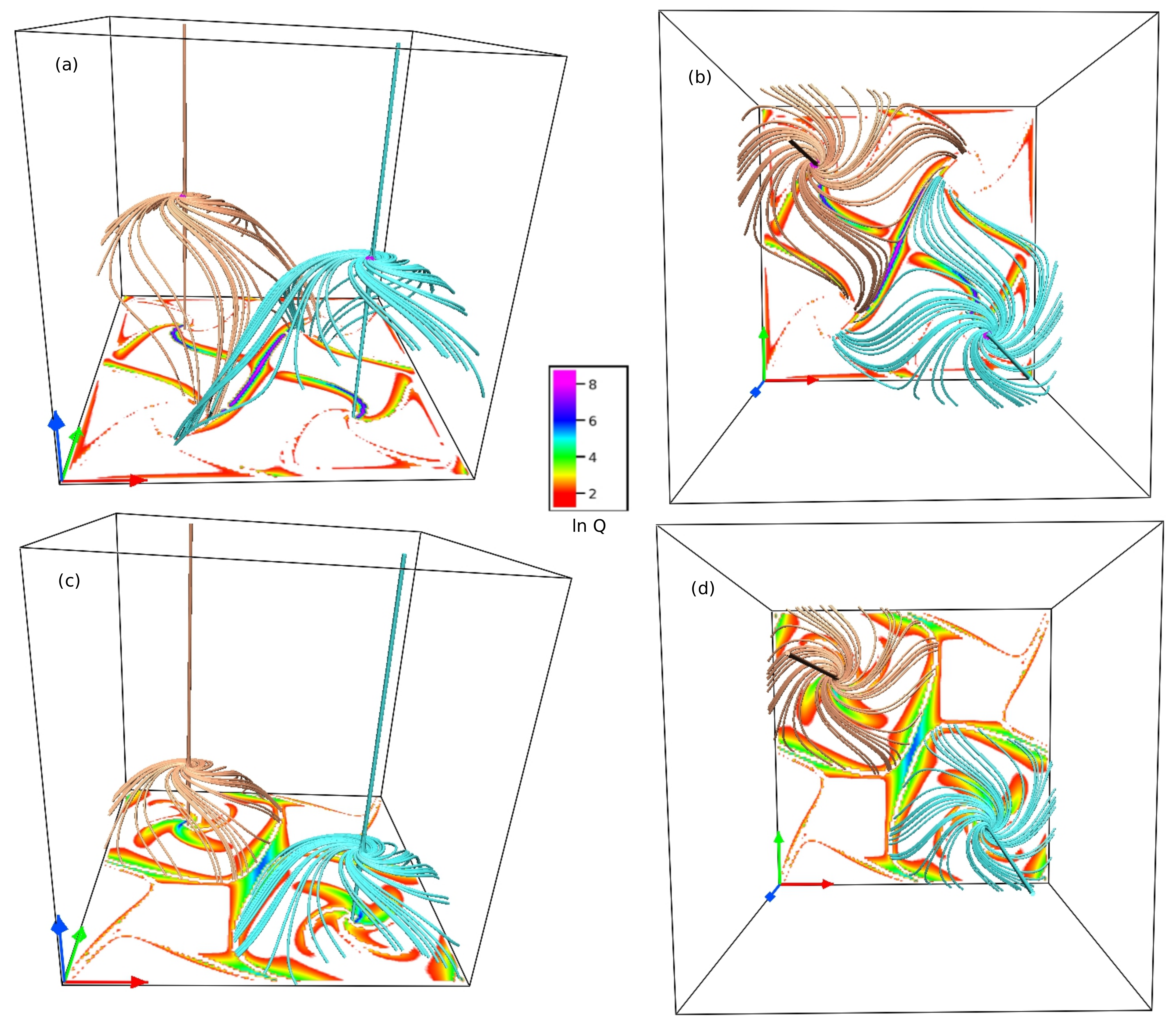}
\caption{The structural skeleton (i.e., the separatrix skeleton plus the QSL quasi-skeleton) of ${\bf{B}}$ with the bottom boundary being superimposed with $\ln Q$ for $c_0=0.1$ (side and top views in panels (a) and (b)) and $c_0=0.5$ (side and top views in panels (c) and (d)). The existence of large $Q$ (such that $\ln Q \in \{2,8 \} $) represents the location of separatrices or QSLs. For both $c_0$ values, notable is the presence of both the separatrices of the coronal nulls and also the QSLs associated with the central quasi-separator at $(\pi, \pi, z)$ and the extra quasi-separators on the boundary. Also interesting is the larger $Q$ values for $c_0=0.1$ than $0.5$ at the central quasi-separator at $(\pi, \pi, z)$. } \label{newfigure4}
\end{figure}

To describe the topological structure of ${\bf{B}}$, in Figure \ref{newfigure3}, we examine its magnetic skeleton by plotting  magnetic nulls, separatrix surfaces and spines \citep{2004A&A...428..595P, 2007RSPSA.463.1097H}. The skeleton of ${\bf{B}}$ is shown for the chosen $c_0=0.1$ (panels (a) and (b)) and $c_0=0.5$ (panels (c) and (d)). For $c_0=0.1$, the panels (a) and (b) of the figure confirm the presence of two 3D nulls located at the height $z\approx 0.955\pi$ over the sites of the two spiral nulls (i.e., at $(\pi/2, 3\pi/2)$ and $(3\pi/2, \pi/2)$) of the transverse field. The 3D nulls have well-defined spine axes and dome-shaped separatrix or fan surfaces whose feet coincide with many of the regions of strong $Q$ in Figure \ref{newfigure4}. Similarly, for $c_0=0.5$, panels (c) and (d) of Figure \ref{newfigure3} show the existence of a pair of 3D nulls over the two spiral nulls of the transverse field. The coordinates of the nulls are $(x,y,z) \approx (\pi/2, 3\pi/2, 0.44\pi)$ and  $(3\pi/2, \pi/2, 0.44\pi)$.
We have also analytically verified the locations of the nulls in $\bf{B}$ which are $(x,y,z) = (\pi/2, 3\pi/2, \ln(2/c_0))$ and $(3\pi/2, \pi/2, \ln(2/c_0))$ for $c_0=0.1$ and $0.5$ --- matching well with the locations obtained from the used numerical technique. Importantly, for both cases, the overall morphology of the nulls is similar to the 3D nulls obtained with the extrapolated coronal fields {\citep{2009SoPh..254...51L, platten-etal2014, prasad+2018apj, 2019ApJ...875...10N, prasad-2020}}. The MFLs constituting the dome-shaped separatrix surfaces predominately intersect the bottom boundary and the intersection points, or the footpoints, trace nearly-closed circular curves --- further advocating the similarity.

Additionally, as shown in Figure \ref{newfigure3}, the 3D null are located at higher heights and, therefore, their separatrix domes are larger for $c_0=0.1$ than $0.5$. Figure \ref{newfigure3}(b) demonstrates that, when viewed from the top, the separatrix surfaces of the 3D nulls for $c_0=0.1$ seem to touch in the vicinity of the points $(x,y)=(\pi, \pi)$ along $z$ and, as a result, the geometry of the MFLs in the vicinity is what is expected for a quasi-separator (or hyperbolic flux tube). The neutral X-line $(\pi, \pi, z)$ of ${\bf{B}}_1$ (see Fig. \ref{newfigure2}(b)) turns into the quasi-separator when the vertical field ${\bf{B}_2}$ is added. To further confirm, we expand components of  ${\bf{B}}$ in a Taylor series in the immediate vicinity of $x=\pi, y=\pi, z=0$ to get 

\begin{eqnarray}
\label{taylor2}
& & B_x=x-y ,\\
& & B_y=-(x+y)+2\pi,\\
& & B_z=c_0, 
\end{eqnarray} 

\noindent which attest the absence of the X-type null at $(x,y,z)=(\pi, \pi, 0)$. However, for both $c_0=0.5$ and $0.1$ the nearby geometry is that of a hyperbolic flux tube.  Interesting is the orientation of the MFLs of the hyperbolic flux tube for $c_0=0.1$, which is expected to be favorable for initiating MRs. In comparison to $c_0=0.1$ case, the magnitude of $\bf{B_2}$ is larger for $c_0=0.5$ case and, consequently, the corresponding separatrix dome surfaces are separate --- leading to the elimination of the favorable orientation in this case (see Fig. \ref{newfigure3}(d)).

For further investigation, in Figure \ref{newfigure4} we illustrate the topological skeleton (3D nulls, corresponding separatrix surfaces and spines) of $\bf{B}$ overlaid with the $Q$-map at the bottom boundary for both $c_0=0.1$ (panels (a) and (b)) and $c_0=0.5$ (panels (c) and (d)). Note that for both the cases, large $Q$ exists at the central region near the point  $(\pi,\pi,0)$. The presence of the large $Q$ suggests that a X-line (i.e., a line of X-points in $z$-constant planes)
located along $(\pi,\pi,z)$ (see Fig.~\ref{newfigure2}) converts into a quasi-separator (or hyperbolic flux tube) by the addition of the constant vertical field $c_0\bf{B}_2$ to $\bf{B}_1$. Noticeably, because of a smaller $c_0$, the $Q$ values in the central region are higher for $c_0=0.1$ in comparison to $c_0=0.5$ --- indicating
a more favorable location for reconnection provided the flows are appropriate. Moreover, the large Q values near the boundaries for $\bf{B}$ (Fig.~\ref{newfigure4}) are almost co-spatial with the rest of the X-type neutral lines of $\bf{B}_1$ (and the X-type neutral points of the transverse field) located at boundaries (Fig. \ref{newfigure2}) --- indicating toward the transformation of all the neutral lines into QSLs. In addition, in the initial field $\bf{B}$, unlike $\bf{B}_1$, QSLs seem to be absent over the two spiral nulls of the transverse field located at $(\pi/2, \pi/2, z)$ and $(3\pi/2, 3\pi/2, z)$. The absence can be attributed to the addition of the constant field ${\bf{B}}_2$ to $\bf{B}_1$. From Figures \ref{newfigure4}(a) and (b), we also note the existence of large $Q$ values near the footpoints of the MFLs of the separatrix dome surfaces for $c_0=0.1$. Similarly,  Figures \ref{newfigure4}(c) and (d) show the presence of the contours of large $Q$ approximately co-located with inner and outer vicinity of the foot-points of the dome separatrices for the case $c_0=0.5$. Relevantly, \citet{2007ApJ...660..863T} suggested that the QSLs determine the quasi-skeleton of a magnetic field and one can define the structural skeleton which is the sum of the topological skeleton and the quasi-skeleton. Hence, Figure \ref{newfigure4} plots the structural skeleton of the initial field ${\bf{B}}$.

Based on the above analysis, overall, the selected initial magnetic fields can be divided into two broad categories. The first one (belonging to $c_0=0.5$) supports a relatively simpler topology with a pair of coronal 3D nulls located at low heights and a central quasi-separator. The corresponding separatrix surfaces do not  touch to each-other and, hence, are independent.  While the second one (corresponding to $c_0=0.1$) also contains a pair of 3D nulls and a central quasi-separator. But, for this case, the coronal nulls are situated at greater heights and the separatrix surfaces appear to interact with the larger $Q$ values in the central region of the computational domain --- making the case more suitable for QSL reconnection in addition to the null point reconnection at the coronal nulls. This further justifies the selection of the two particular $c_0$ values.

\section{Governing MHD Equations and Numerical Model} 
\label{num-model}

The presented simulations are carried out by numerically solving the incompressible Navier-Stokes MHD equations under the assumption of thermal homogeneity and perfect electrical conductivity  \citep{sanjay2016}. The MHD equations in dimensionless form are: 

\begin{eqnarray}
\label{stokes}
& & \frac{\partial{\bf{v}}}{\partial t} 
+ \left({\bf{v}}\cdot\nabla \right) {\bf{ v}} =-\nabla p
+\left(\nabla\times{\bf{B}}\right) \times{\bf{B}}+\frac{\tau_a}{\tau_\nu}\nabla^2{\bf{v}},\\  
\label{incompress1}
& & \nabla\cdot{\bf{v}}=0, \\
\label{induction}
& & \frac{\partial{\bf{B}}}{\partial t}=\nabla\times({\bf{v}}\times{\bf{B}}), \\
\label{solenoid}
 & & \nabla\cdot{\bf{B}}=0, 
\label{e:mhd}
\end{eqnarray}
written in usual notations. The various variables in the MHD equations are normalized as follows 

\begin{equation}
\label{norm}
{\bf{B}}\longrightarrow \frac{{\bf{B}}}{B_0},\quad{\bf{v}}\longrightarrow \frac{\bf{v}}{v_a},\quad
 L \longrightarrow \frac{L}{L_0},\quad t \longrightarrow \frac{t}{\tau_a},\quad
 p  \longrightarrow \frac{p}{\rho {v_a}^2}. 
\end{equation}

\noindent The constants $B_0$ and $L_0$ are generally arbitrary, but can be fixed using the average magnetic field strength and size of the system. Here, $v_a \equiv B_0/\sqrt{4\pi\rho_0}$ is the Alfv\'{e}n speed and $\rho_0$ is the constant mass density. The constants $\tau_a$ and $\tau_\nu$ represent the Alfv\'{e}nic transit time ($\tau_a=L_0/v_a$) and viscous dissipation time scale ($\tau_\nu= L_0^2/\nu$), respectively, with $\nu$ being the kinematic viscosity. Notably, the choice of incompressibility (Equation \ref{incompress1}) leads to the volume preserving flow --- an assumption routinely used in other works \citep{dahlburg+1991apj,aulanier+2005aa}. While compressibility plays an important role in the thermodynamics of
coronal loops \citep{ruderman&roberts2002apj}, in this work, our focus is on the changes in magnetic topology idealized with a thermally homogeneous magnetofluid. Utilizing the discretized incompressibility constraint, the pressure perturbation, denoted by $p$,  satisfies an elliptic boundary value problem on the discrete integral form of the momentum equation (Equation \ref{stokes}); cf. \citet{bhattacharyya+2010phpl} and the references therein.

To numerically solve the MHD equations  (\ref{stokes})-(\ref{solenoid}), we note that 
a MHD based numerical model aiming to simulate solar corona must accurately preserve the flux-freezing by minimizing numerical dissipation and dispersion errors away
from the reconnection regions characterized by steep gradients of the magnetic field {\citep{bhattacharyya+2010phpl}}. Such minimization is a signature of a class of inherently nonlinear high-resolution transport methods that preserve field extrema along flow trajectories, while ensuring
higher-order accuracy away from steep gradients in advected fields. Consequently, we utilize the well established magnetohydrodynamic numerical model EULAG-MHD \citep{smolarkiewicz&charbonneau2013jcoph}. The model is an extension of the hydrodynamic model EULAG predominantly used in atmospheric and climate research \citep{prusa2008cf}. Here we discuss only essential features of the EULAG-MHD and refer the readers to \citet{smolarkiewicz&charbonneau2013jcoph} and references therein for detailed discussions. The model is based on the spatio-temporally second-order accurate non-oscillatory forward-in-time multidimensional positive definite advection transport algorithm, MPDATA \citep{smolarkiewicz2006ijnmf}.  
Importantly, MPDATA has the proven dissipative property which, intermittently and adaptively, regularizes the under-resolved scales by simulating magnetic reconnections 
and mimicking the action of explicit subgrid-scale turbulence models \citep{2006JTurb...7...15M} in the spirit of
Implicit Large Eddy Simulations (ILES)
\citep{grinstein2007book}. Arguably, the residual numerical dissipation is then negligible 
everywhere but at the sites of MRs. Moreover, this dissipation being intermittent in time and space, a quantification of it is meaningful only in
the spectral space where, analogous to the eddy viscosity of
explicit subgrid-scale models for turbulent flows, it only acts on the shortest modes admissible on the grid; in particular, in the vicinity of steep gradients in
simulated fields. Such ILESs conducted with the model have already been successfully utilized to simulate reconnections to understand their role in the coronal dynamics \citep{prasad+2017apj,prasad+2018apj,2019ApJ...875...10N}. In this work, the presented computations continue to rely on the effectiveness of ILES in regularizing the under-resolved scales by commencement of magnetic reconnections.

\section{Simulation Results}
\label{results}
The simulations are carried out on a grid of uniform resolution $128\times 128\times 128$, resolving the domain
$\Gamma$. The initial states are characterized by the magnetic field ${\bf{B}}$ given by equations (\ref{bx})-(\ref{bz}) and the velocity field ${\bf{v}}=0$. Simulations are performed with $c_0 = 0.1$ and $0.5$. The lateral boundaries ($x$ and $y$) are chosen to be periodic, while magnetic fluxes at the vertical boundaries are kept fixed to zero {\citep{dinesh2015}}. At the
bottom boundary, the $z$-components of ${\bf{B}}$ and  ${\bf{v}}$ are kept fixed to their initial values (line-tied boundary condition). In the conducted simulations, the dimensionless coefficient $\tau_a$/$\tau_\nu \approx 10^{-4}$, which is roughly one orders of magnitude larger than its coronal value \citep{prasad+2018apj}. The larger $\tau_a$/$\tau_\nu$, however, is expected to only speed-up the evolution  without an effect on the corresponding change in the topology of MFLs. The initial Lorentz force pushes the plasma from the initial static state and imparts dynamics. To examine the onset of MRs, in the following, we analyze the evolution of the two cases $c_0=0.1$ and $0.5$ separately. For $c_0=0.5$, the 3D nulls are located at lower heights and the corresponding separatrix domes are fairly independent (see Fig. \ref{newfigure3}). Therefore, we first consider this case.

\subsection{Case (I) $c_0=0.5$ }
\begin{figure}[htp]
\centering
\includegraphics[angle=0,scale=.24]{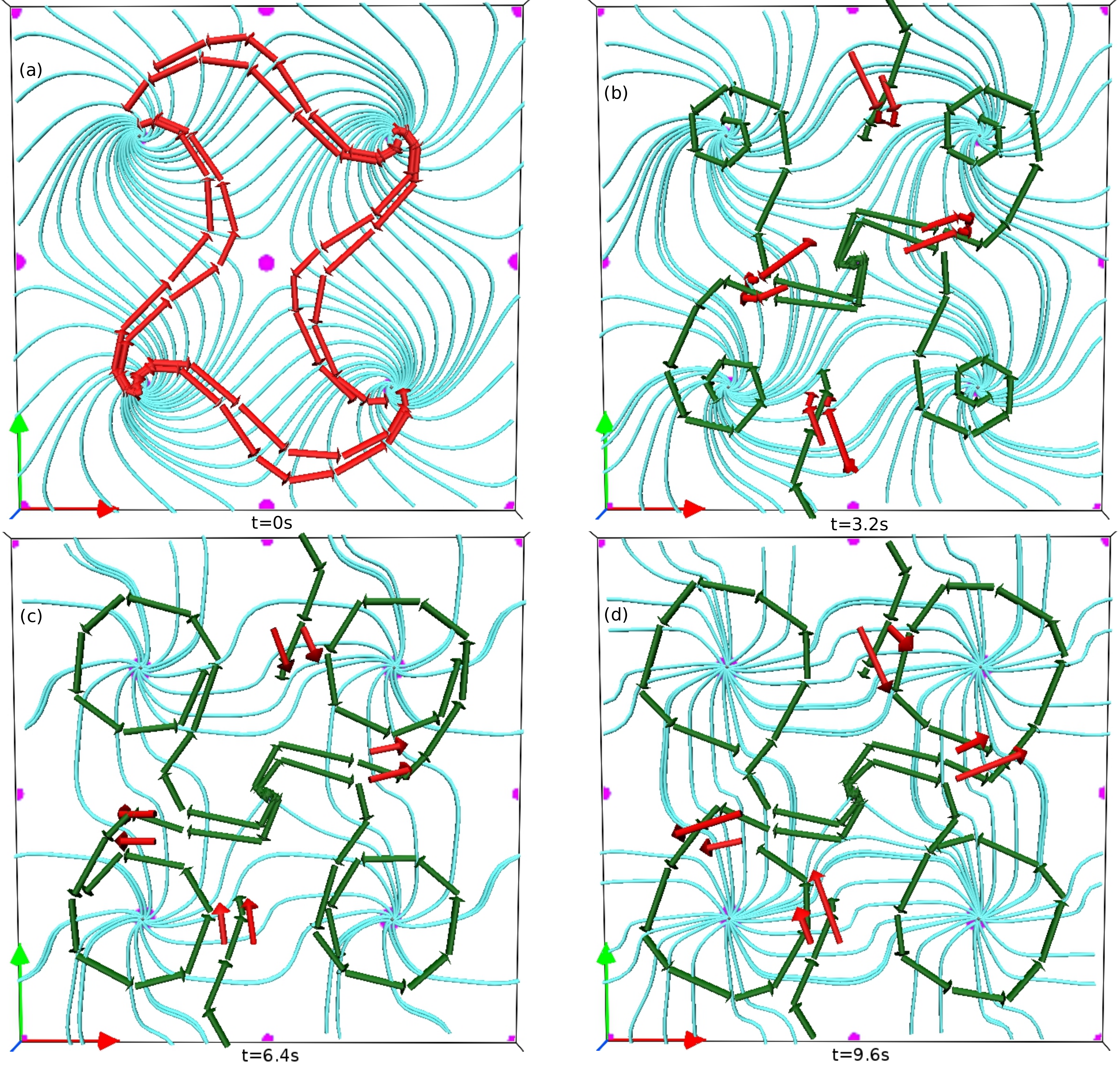}
\caption{ Evolution of field lines of the transverse field (in cyan color) at $z=0$ plane for $c_0=0.5$. The figure is further overplotted with the streamlines of the flow (in green color), Lorentz force (red arrows) and, the neutral points (in pink). 
   } \label{newfigure5}
\end{figure}

This case belongs to the initial magnetic field which includes a pair of 3D nulls and a central quasi-separator. For a careful inspection of the simulated dynamics, in Figure \ref{newfigure5}, we first present the evolution of the transverse field overlaid with the plasma flow (projected at $z=0$ plane) and the Lorentz force at the bottom boundary. Notable are the reversal of the direction of initial Lorentz force (marked by red color) and the generation of rotational flow (in green color) around the spiral nulls in the early phase of evolution. 

\begin{figure}[htp]
\centering
\includegraphics[angle=0,scale=.23]{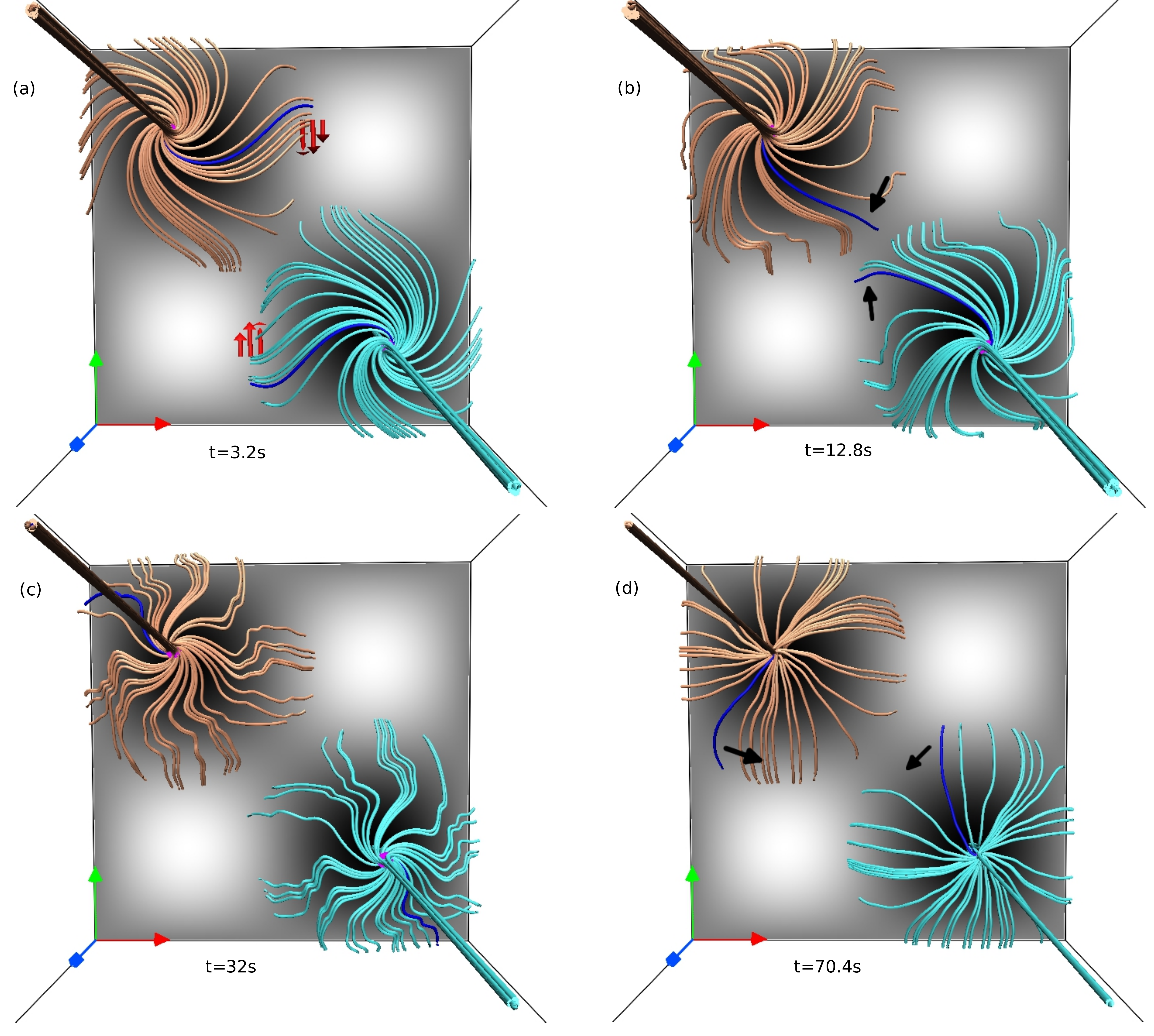}
\caption{  Evolution of the topological skeleton of $\bf{B}$ for $c_0=0.5$ (Figs. \ref{newfigure2}(c) and (d)). Panel (a) is overlaid with Lorentz force (red arrows) at the bottom boundary.  The rotation of the separatrix surfaces is evident from the movement of the blue MFLs (further marked by black arrows).     } \label{newfigure6}
\end{figure}

Figure \ref{newfigure6} shows the time sequences of the magnetic skeleton of the initial field  $\bf{B}$ (Fig. \ref{newfigure3}). For plotting the evolution of the MFLs, in this and the subsequent figures, we utilize ``field line advection" technique in-built in the VAPOR visualization package \citep{clyne-2005} in which one
representative point for a selected MFL is advected by the velocity field and, then the advected point is used as a seed to plot the MFL at later time \citep{mininni-2008}. For a detailed description of the technique and its successful illustration in ideal as well as non-ideal magnetofluids, readers are referred to \citet{clyne-2007, mininni-2008}.
Noticeably, under the ideal conditions, the technique is similar to other methods of tracking of MFLs in which one follows the motion of individual plasma elements and traces the changes in the MFLs that are attached to those plasma elements \citep{2003ApJ...595.1259L}.    
From Figure \ref{newfigure6}, notable is the rotation of the separatrix domes of the 3D nulls. In the figure, black arrows and motion of the blue MFLs clearly mark the direction of the rotation.  The rotation appears to be initiated by the Lorentz force after $t=3.2s$ (see Figs. {\ref{newfigure5}} and \ref{newfigure6}(a)). Initially, the rotation is in clockwise direction 
(Fig.~\ref{newfigure6}(b)). This increases the twist and hence, the tension in the MFLs of the separatrix domes (panel (c)). Eventually, the magnetic tension changes the direction of rotation and the MFLs rotate in counter-clockwise direction (panel (d)).  The twist of the MFLs then decreases with time.
 Such twisting and untwisting rotational motion of the MFLs is expected to repeat in time and, ultimately, get damped by the viscous drag force.

\begin{figure}[htp]
\centering
\includegraphics[angle=0,scale=.21]{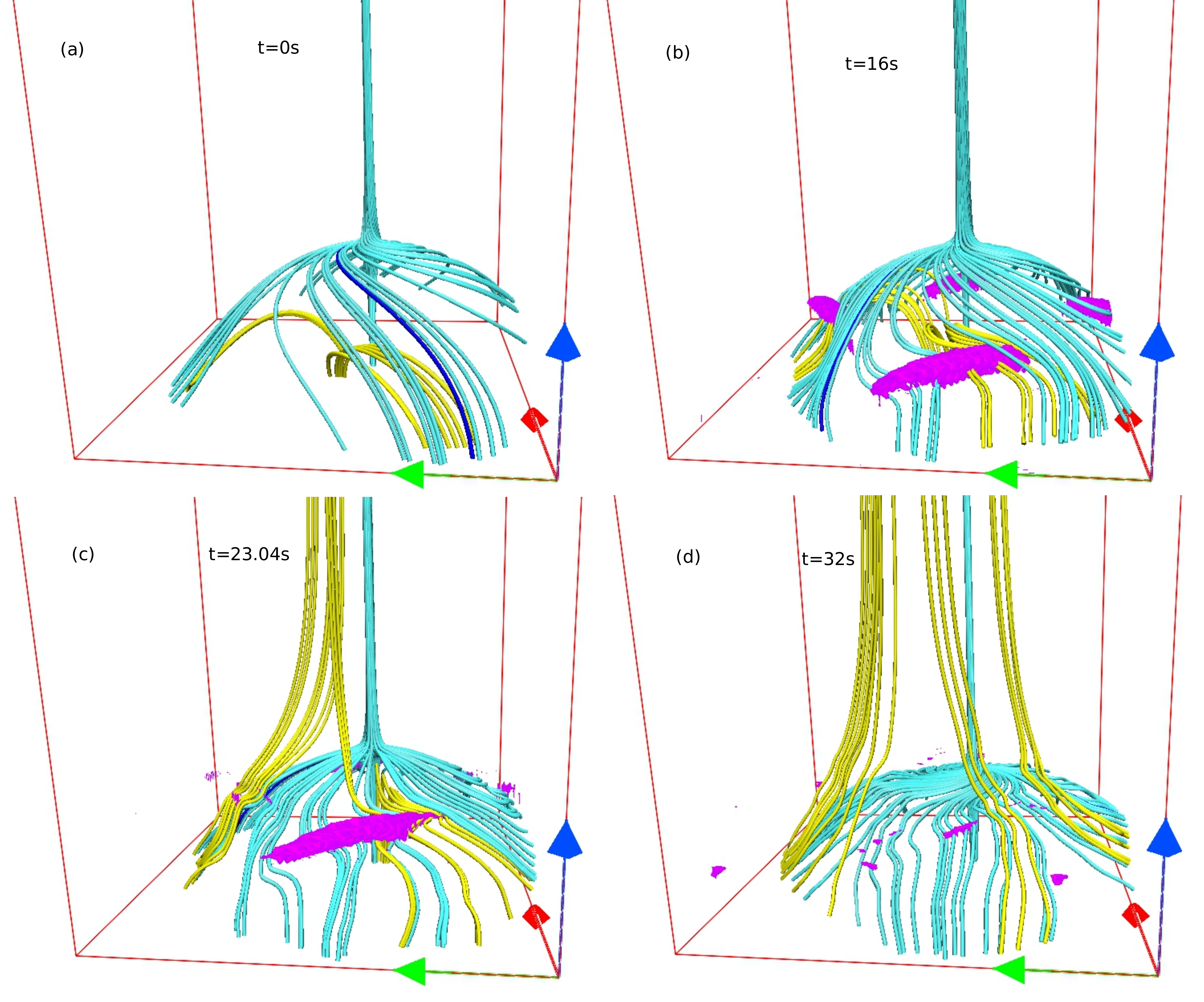}
\caption{ Evolution of a 3D null along with the fan surface (represented by cyan MFLs) for $c_0=0.5$. The figure is further overplotted with two set of MFLs (in yellow) situated below the dome and the $J-70$ surfaces (in pink). The movement of MFLs of the dome is marked by a blue MFL.  Important are the appearances of the $J-70$ surfaces at the fan surface and change in the connectivities of the yellow MFLs.     
} \label{newfigure7}
\end{figure}

To investigate the MRs at the 3D nulls, the time evolution of the spine and the separatrix fan surface of a 3D null located at $(3\pi/2, \pi/2, 0.44\pi)$ is shown in Figure \ref{newfigure7}. In the figure, we also plot two set of MFLs (in yellow) which are situated under the dome-shaped fan surface at $t=0s$. Moreover, to demonstrate the current sheet formation, we overlay the figure with isosurfaces of current density $\mid{\bf{J}}\mid$ having an isovalue which is $70\%$ of the maximal value of $\mid{\bf{J}}\mid$. 
The selection of the isovalue is based on an optimization of  constructing a smooth and identifiable isosurface with a large isovalue. 
We call these isosurfaces $J-70$ and identify them as the CSs because they are 2D manifolds and not the boundaries of 3D volumes. Notably, the yellow MFLs do not appear to co-rotate with the MFLs of the fan surface (as evident from the motion of the blue MFL). This seems to generate favorable contortion in the MFLs --- making the yellow MFLs and the MFLs of the fan, non-parallel. Consequently, the CSs develop in the vicinity of the fan surface (panel (b)). In addition, the yellow MFLs rise toward the 3D null and, eventually come out of the dome.  This is a clear indication of the change in
connectivities of the yellow MFLs, suggesting the occurrence of torsional fan reconnections at the 3D null \citep{2009PhPl...16l2101P,  pontin-2013}. With reconnections, the CSs dissipate and the contortion in the MFLs decrease with time (panel (d)). Similar evolution is found for the other 3D null (not shown).

\begin{figure}[htp]
\centering
\includegraphics[angle=0,scale=.21]{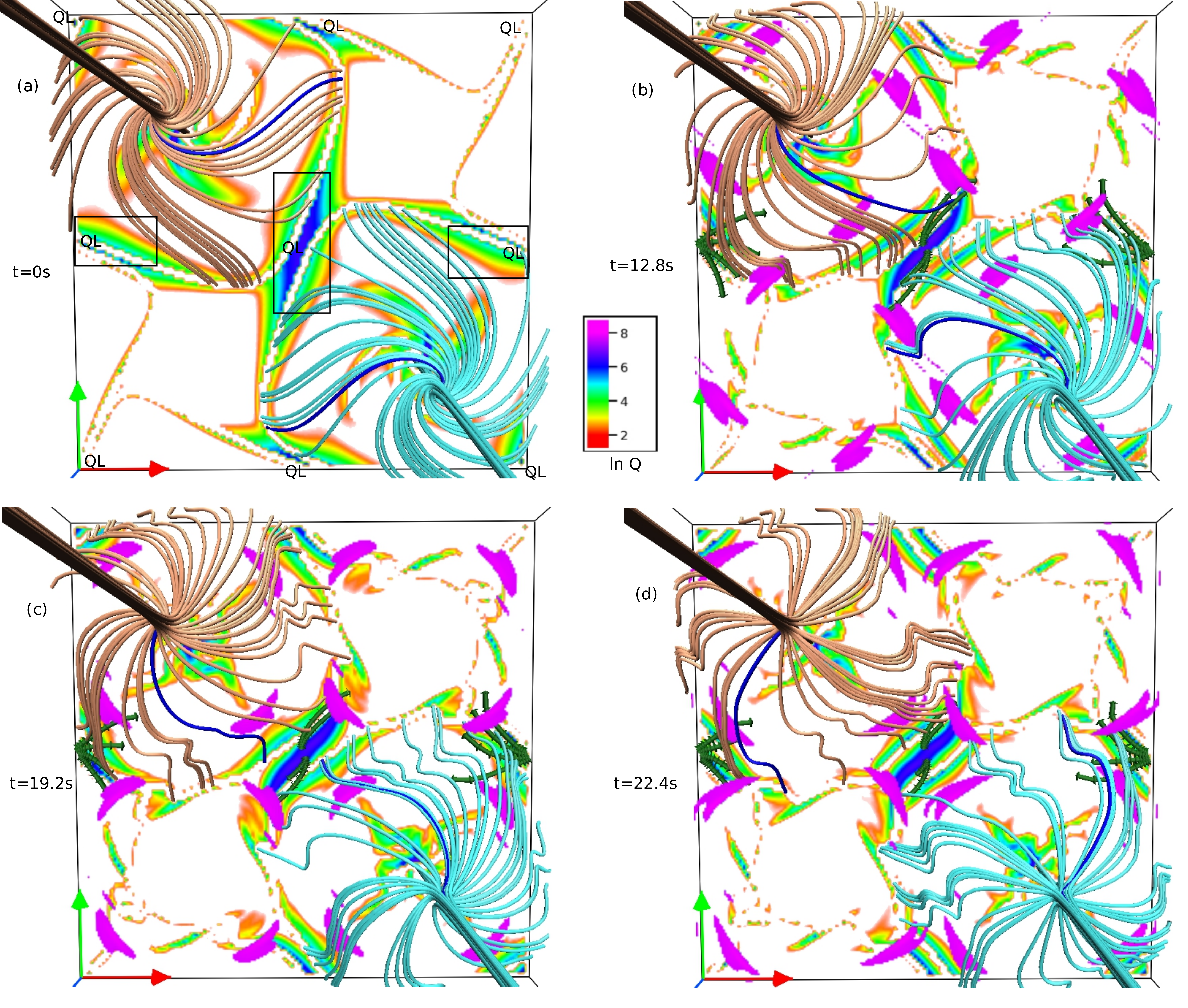}
\caption{ Evolution of 3D nulls with bottom boundary being overplotted with $ln Q$ ($c_0=0.5$). The regions of strong $Q$ in the initial field are marked by QL in 
panel (a).  We also plot the streamlines of plasma flow (green
arrows) near the three QSLs marked by rectangular boxes in panel (a), some of which are separatrices. Currents having sufficiently high values are shown at a $z$-constant plane (in pink). The motion of the dome MFLs is depicted from
the blue MFLs. The direction of the MFLs movement is largely different from the flow direction --- manifesting the flipping MFLs in the separatrices and QSLs.   
} \label{newfigure8}
\end{figure}

Figure \ref{newfigure8} depicts the time profile of the $Q$-map at the bottom boundary overlaid with the 3D nulls. To locate the QSLs, we plot the skeleton of the separatrix surfaces and, then the extra features in the Q-map (marked by QL in Fig. \ref{newfigure8}(a)) are identified as the QSLs (also shown in Fig. \ref{newfigure4}). Relevantly, as mentioned in Section 2, these QSLs correspond to the X-type nulls of the transverse field (see Figs.~\ref{newfigure2} and \ref{newfigure5}). To explore the possibility of reconnections near the QSLs, Figure \ref{newfigure8} is overlaid with plasma flow (green arrows) near the regions of three QSLs marked by rectangular boxes in Figure \ref{newfigure8}(a), as representative cases. Additionally, the current density $\mid{\bf{J}}\mid$ having values around $35\%$ of its maximal value are plotted on a $z$-constant plane (in pink) situated near the bottom boundary. Notably, the direction of plasma flow (green arrows) is visibly different from the direction of the motion of MFLs (showcased by the blue MFLs) in the vicinity of the separatrices --- a telltale sign of the flipping or slipping of field lines \citep{1992JGR....97.1521P, 2006SoPh..238..347A, 2017JPlPh..83a5301J}. Moreover, with time, the currents start to appear near the fan separatrix regions (although remains negligible at $(\pi, \pi, 0)$) --- further supporting the CS development and the onset of the reconnections in the vicinity of the separatrices. Under the simulated viscous relaxation, such appearances of the CSs can be attributed to the autonomous development of the favorable forcing {\citep{sanjay-2015}}. However, the strength of the currents near QSLs is almost half of the strength at the fan surfaces of the 3D nulls (Fig. \ref{newfigure7}) --- indicating  the reconnections near QSLs to be less energetically efficient than the 3D nulls. This supports  the  proposal  of  \citet{priest-forbes1989,1992JGR....97.1521P} that the more efficient reconnections require the  favorable  geometry of MLFs (such as separatrix,  QSL,  null  or  separator) as well as the favorable flows. It appears that both are present at the 3D nulls, while the  favorable flow is missing in the case of the QSL around the point $(\pi, \pi, 0)$.

\subsection{Case (II) $c_0=0.1$ }
\begin{figure}[htp]
\centering
\includegraphics[angle=0,scale=.24]{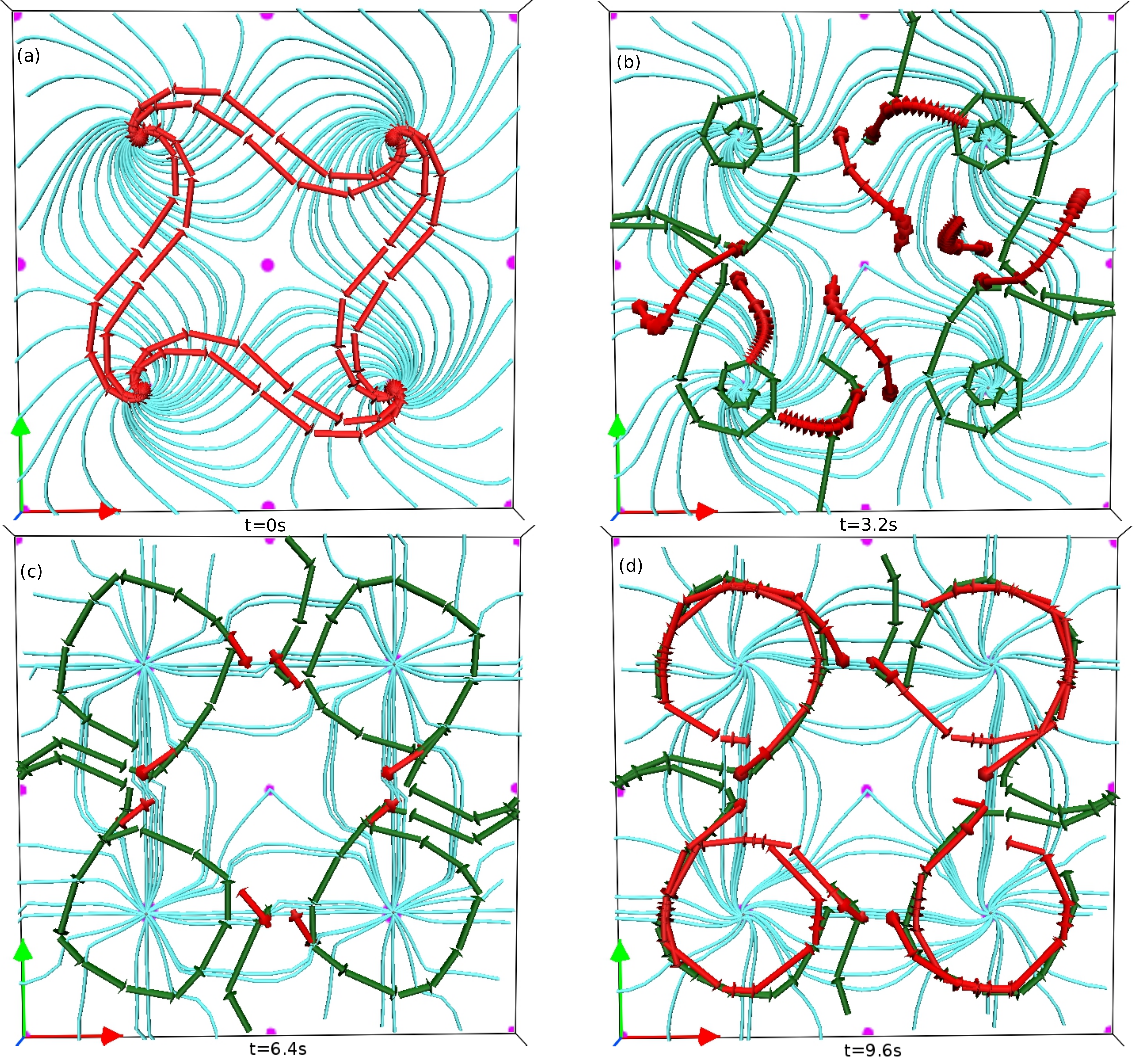}
\caption{ Time sequences of the transverse field (in cyan color) at the bottom boundary for $c_0=0.1$. The flow (green arrows), the Lorentz force (red arrows) and the neutral points (in pink) are also overplotted in the figure.   } \label{newfigure9}
\end{figure}

As found in the initial field, the 3D nulls for this case are located at greater heights in comparison to $c_0= 0.5$ and the separatrix surfaces touch in the central region that is located around the line $(\pi, \pi, z)$ --- leading to an MFL geometry favorable to MRs.
Figure \ref{newfigure9} depicts the time sequences of the field lines of the transverse field during their 
evolution. The figure also  plots plasma flow (denoted by green arrows) projected on the lower boundary and the Lorentz force (marked by red arrows). Noticeably, in response to the initial Lorentz force, a rotational flow is produced near the spiral nulls of the  transverse field.

\begin{figure}[htp]
\centering
\includegraphics[angle=0,scale=.24]{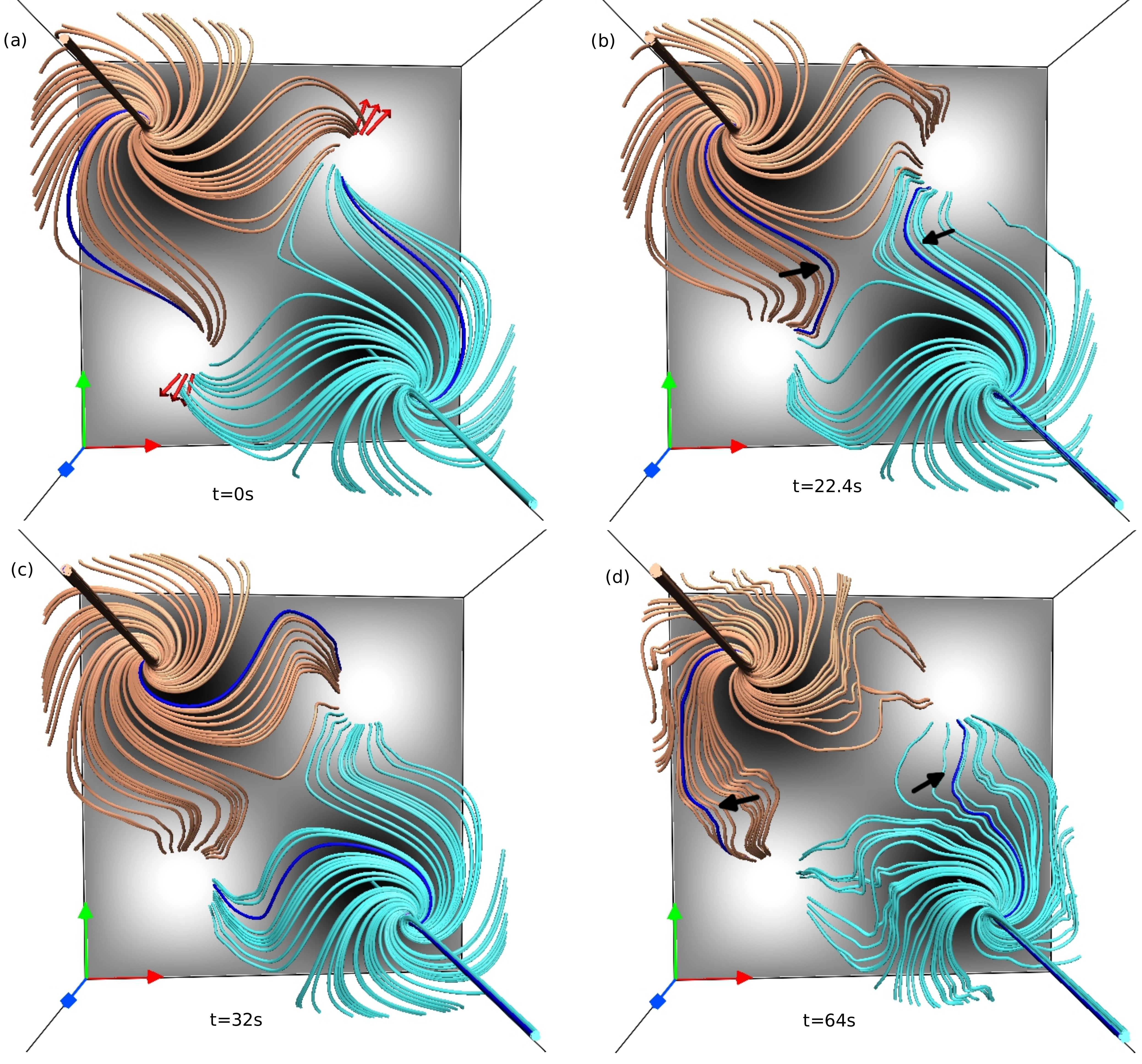}
\caption{ Time profile of the magnetic skeleton (3D nulls, separatrix dome surfaces and spines) of $\bf{B}$ for $c_0=0.1$ (see Figs.~\ref{newfigure2}(a) and (b)). Red arrows in panel (a) show the direction of Lorentz force at the lower boundary. Evident is rotational motion of the separatrix domes, as illustrated by the blue MFLs and black arrows. } \label{newfigure10}
\end{figure}

 To have an overall understanding of the dynamics, in Figure \ref{newfigure10}, we show the time sequences of the topological skeleton in the forms of the 3D nulls and the corresponding spines and separatrix surfaces. The initial Lorentz force (marked by red arrows in panel (a)) appears to push the footpoints of the separatrix domes and initiate the rotational motion of the domes (also evident from Fig. \ref{newfigure9}). When viewed from the top, the rotation is in counter-clockwise direction --- illustrated by the blue
MFLs and black arrows. Similar to  the case of $c_0=0.5$, it enhances the twist and, consequently, magnetic tension in the MFLs
 (see Fig. \ref{newfigure10}(c)) which, finally, reverses the direction of rotation in clockwise direction (cf. Fig. \ref{newfigure10}(d)). The rotational motion is found to oscillate in time and eventually gets decayed by the viscosity.

\begin{figure}[htp]
\centering
\includegraphics[angle=0,scale=.21]{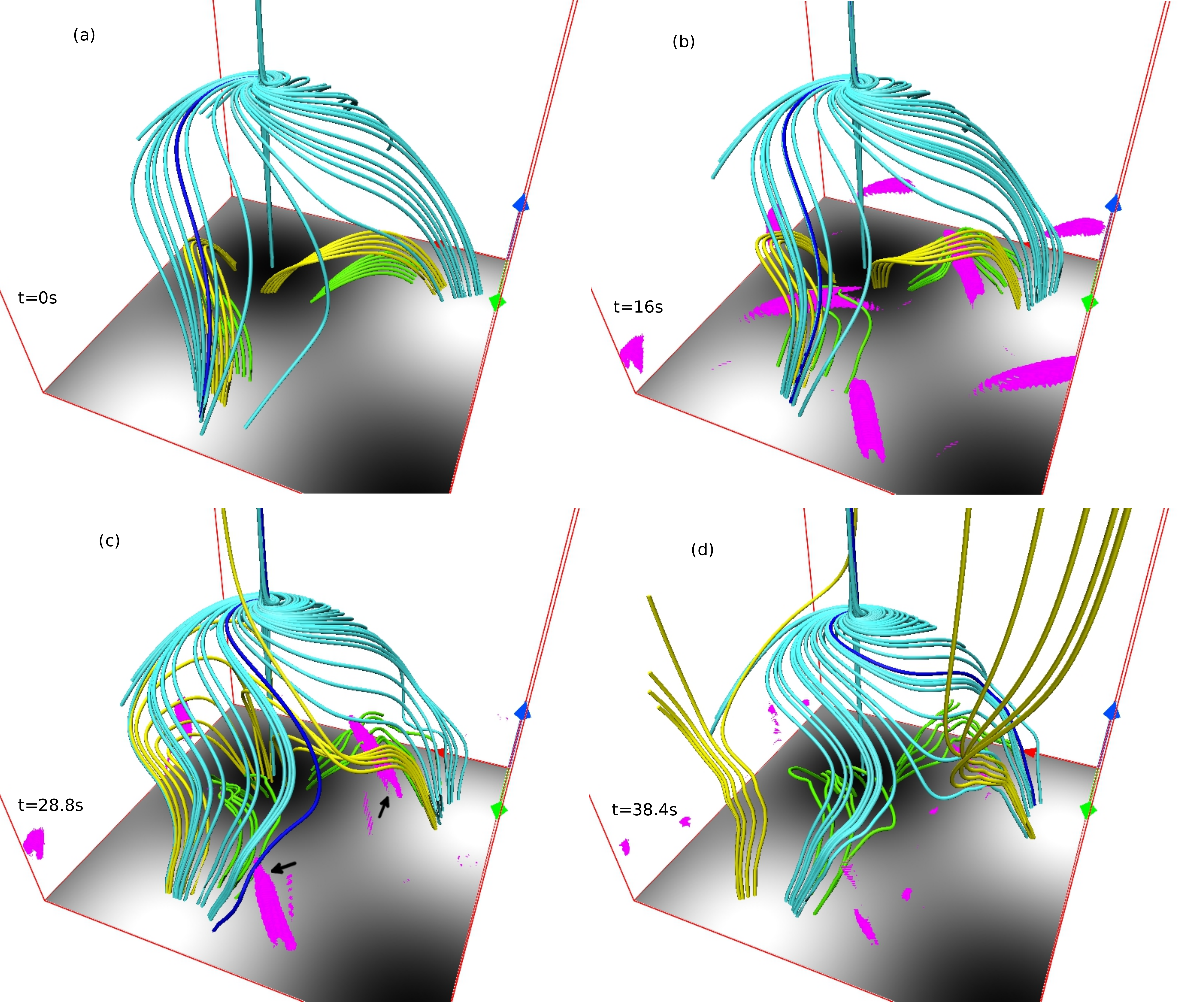}
\caption{  Time evolution of a 3D null with the fan surface and spine (denoted by cyan MFLs) for $c_0=0.1$. The figure is further overlaid with two set of magnetic loops located at different heights (in colors green and yellow) under the fan surface at $t=0s$ and, the $J-70$ surfaces (in pink). The motion of dome MFLs can be tracked by a blue MFL. Notable are the appearances of the $J-70$ surfaces inside the fan surface which later extend toward the fan (marked by black arrows in panel (c)). 
With time, the yellow MFLs change their connectivities.   } \label{newfigure11}
\end{figure}

\begin{figure}[htp]
\centering
\includegraphics[angle=0,scale=.20]{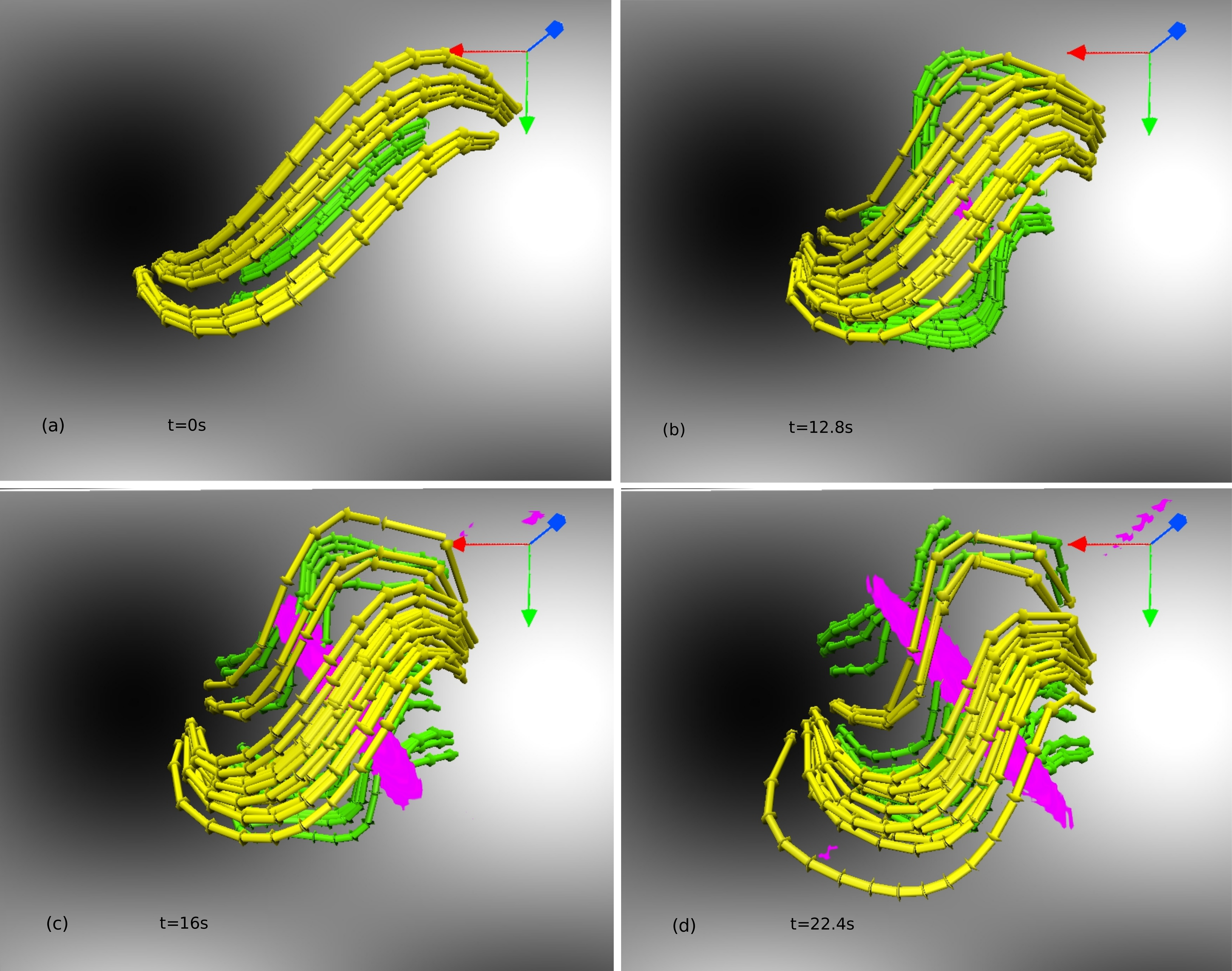}
\caption{  Time evolution of MFLs in the vicinity of a $J-70$ surface for $c_0=0.1$. Initially, the MFLs are in form of closed parallel loops situated at two different heights (panel (a)). As time progresses, the loops become increasingly non-parallel, resulting in the formation of CS.   } \label{newfigure12}
\end{figure}

 To explore the initiation of MRs at the 3D nulls, in Figure \ref{newfigure11}, we display the evolution of a 3D null situated at $(3\pi/2, \pi/2, 0.955 \pi)$ along with the corresponding separatrix surface. 
 The figure is further overlaid with the $J-70$ surfaces and two sets of magnetic loops of different heights, initially located under the separatrix dome.
 Importantly, in this case, the CSs appear to form below the dome surface and, then extend toward the dome (marked by black arrows in Fig. \ref{newfigure11}(c)). The figure indicates that the initially parallel yellow and green loops become increasingly non-parallel and lead to the CS formation. To confirm this, in Figure \ref{newfigure12}, we analyze the evolution of MFLs in the vicinity of a $J-70$ surface. At $t=0$s, the MFLs are in the form of 
two different loop systems situated at two different heights
(Fig. \ref{newfigure12}(a)). The 
corresponding MFLs at lower and higher heights are marked by colors green and yellow respectively. The arrowheads represent the directions of the MFLs. These initially parallel MFLs
start to become non-parallel from $t \approx 10$s onward, ultimately leading to the appearance of 
the $J-70$ surface and its
subsequent spatial extension. Such spontaneous development of CSs is in accordance with the Parker's magnetostatic theorem. Further, from Figure \ref{newfigure11}, as the CSs approach the separatrix surface, the yellow MFLs move toward the 3D null and change their connectivities from the inner to the outer connectivity domain. This reveals the onset of the MRs at the 3D null. Identical dynamics is realized for the other 3D null also (not shown).

\begin{figure}[htp]
\centering
\includegraphics[angle=0,scale=.20]{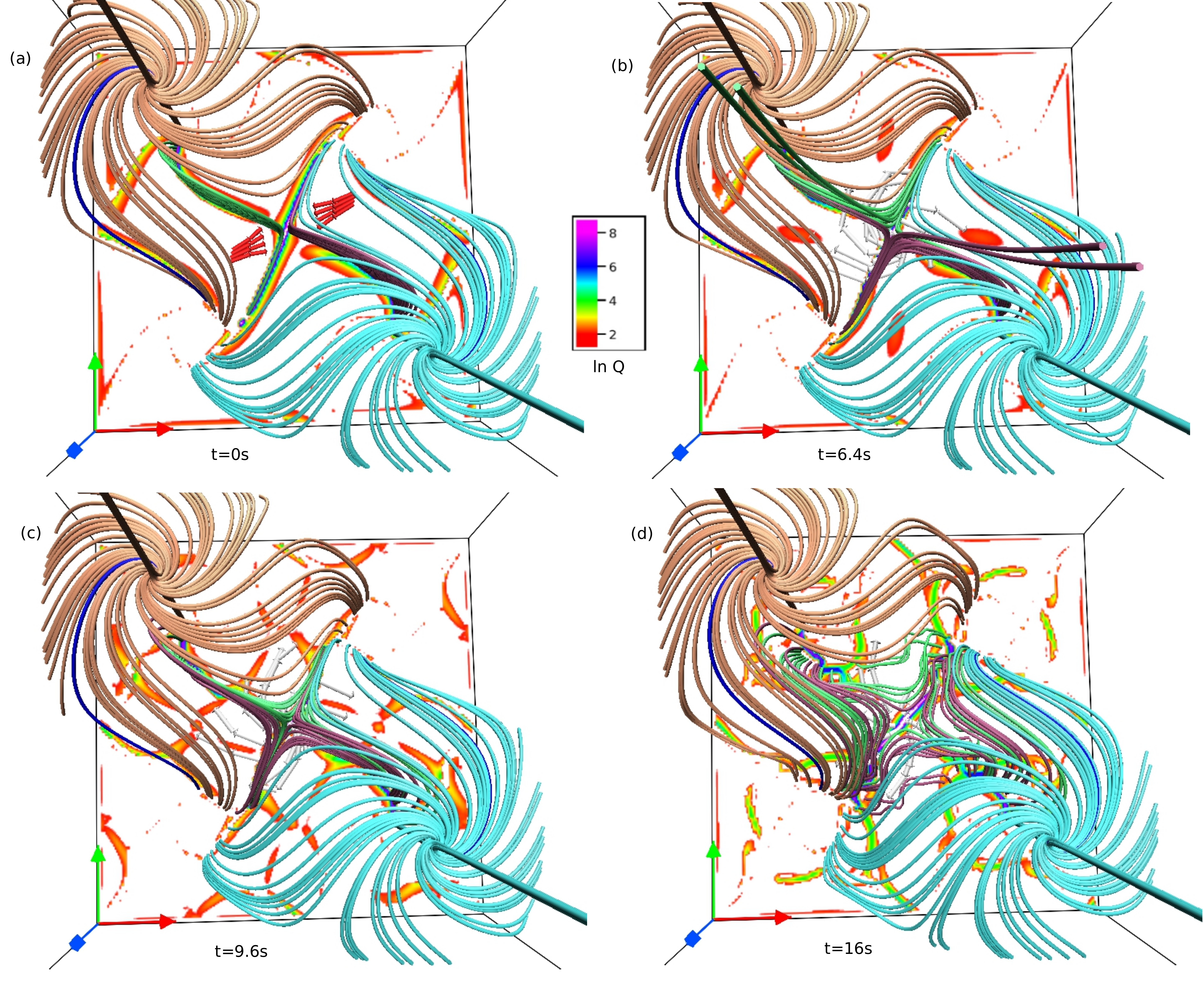}
\caption{Time profile of 3D nulls with $ln Q$ being superimposed at the lower boundary for $c_0=0.1$. Two set of MFLs (in colors purple and green) are plotted near the QSL located in the central region around $(\pi, \pi, z)$. 
The figure also shows the  streamlines of plasma flow (grey arrows) and initial Lorentz force (red arrows) near the QSL. 
Notable is the change in the topology of the purple and green MFLs. } \label{newfigure13}
\end{figure}

Furthermore, to explore the possibility of the reconnections at the QSLs,  we show the time evolution of the topological skeleton superimposed with the $Q$-map at the bottom boundary in Figure \ref{newfigure13}. To keep the presentation tidy, we focus only on the QSL located around the line $(\pi,\pi,z)$ as a representative case. For this case, the domes almost touch each-other and result-in a favorable MFL geometry around the line (Fig.~\ref{newfigure3}). To clearly illustrate this, in Figure \ref{newfigure13}, we further plot two set of MFLs (in colors purple and green) near the line. Moreover, the plasma flow (white arrows), tangential to the $z=0$ plane, is depicted in the vicinity of the QSL location. Notably,  at $t=0s$, the geometry of the purple and green MFLs is what is expected for a quasi-separator or hyperbolic flux tube. Under the favorable initial Lorentz force (marked by red arrows in panel (a)), the oppositely directed purple and green MFLs are pushed toward each other. With time, the MFLs appear to change their connectivities, as evident from the panels (b)-(d) of the figure. This is a marker of  reconnections, which repeat in time, near the QSL. The post-reconnection MFLs move away from central region around the line $(\pi,\pi,z)$ because of the plasma flow (panel (d)). Here also, like $c_0=0.5$, the CSs develop near the QSL location (not shown). For $c_0=0.1$, the rotating sepratrix domes being in close proximity, interact rather strongly about the $(\pi,\pi,z)$ line and cause the reconnections at the QSL that are more prominent in comparison to the case $c_0=0.5$.   
Identical dynamical evolution is observed near the other QSLs located above the X-type nulls of the transverse field (Fig. \ref{newfigure2}(a)) for $c_0=0.1$ which is not presented here.

\begin{figure}[htp]
\centering
\includegraphics[angle=0,scale=.40]{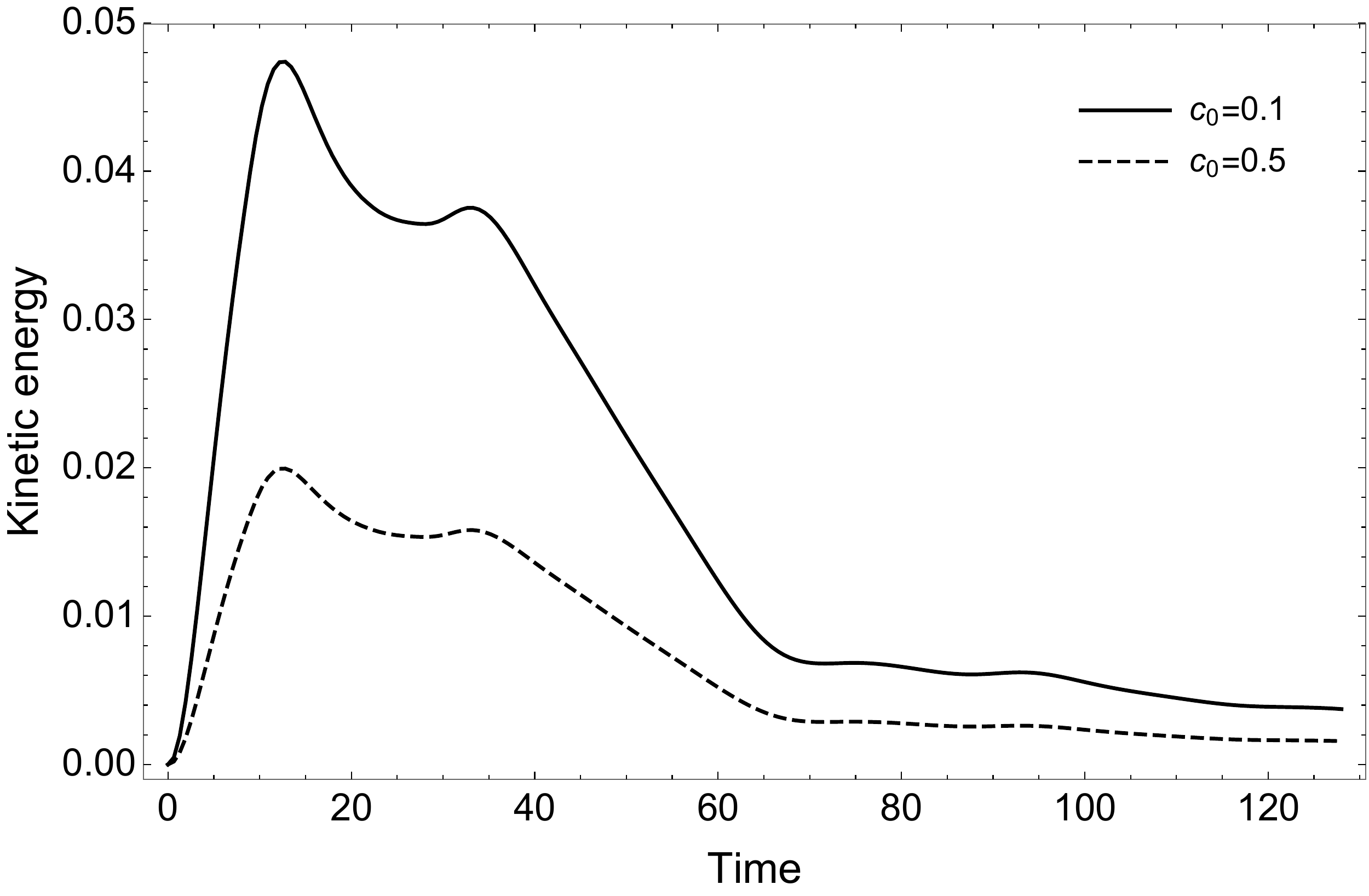}
\includegraphics[angle=0,scale=.40]{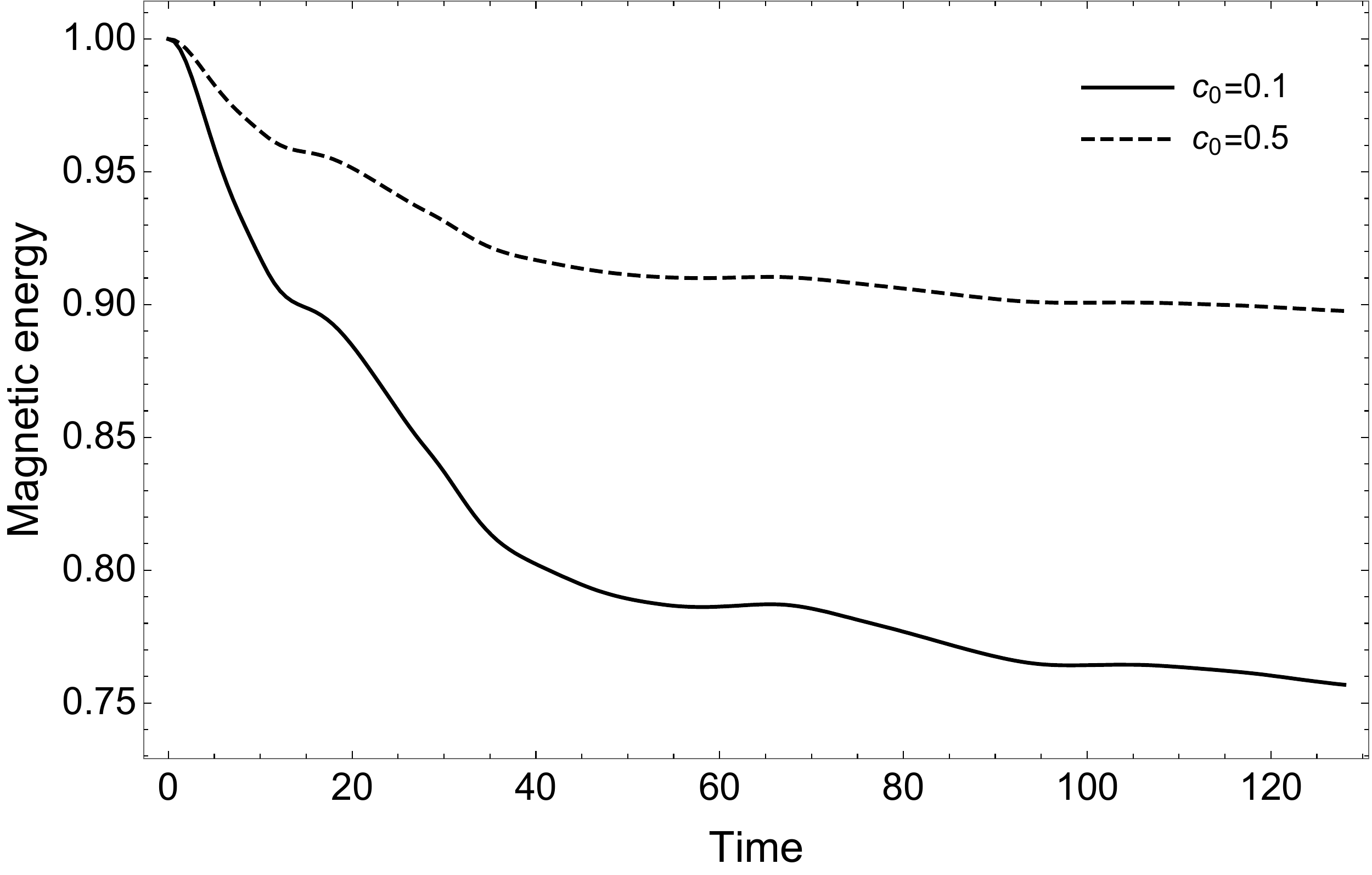}
\caption{ Histories of kinetic (top panel) and magnetic (bottom panel) energies for $c_0=0.1$ and $c_0=0.5$. The solid line is for $c_0=0.1$, while the dashed line corresponds to $c_0=0.5$. The energies are normalized to the initial total energies. Notable is the higher height of the peaks in the kinetic energy and the larger decay of the magnetic energy for $c_0=0.1$ than $c_0=0.5$.
} \label{newfigure14}
\end{figure}

To have an overall comparison of the dynamics for $c_0=0.5$ and $c_0=0.1$, in Figure \ref{newfigure14}, the histories of kinetic and magnetic energies (normalized to the corresponding initial total energies) are plotted for $c_0=0.1$ and $0.5$. For both the cases, the plasma flow is generated via the corresponding initial Lorentz force and the MRs. Subsequently, the flow gets arrested by the viscous drag, leading to the formation of peaks in kinetic energy plots. From Equation \ref{lorentz}, evident is the larger magnitude of the Lorentz force for $c_0=0.5$ than $c_0=0.1$. Although, the height of kinetic
energy peaks for $c_0=0.1$ is greater in comparison to $c_0=0.5$ (top panel of Fig. \ref{newfigure14}). In addition, the depletion of the magnetic energy is larger for $c_0=0.1$ (around $24 \%$ from its initial value) than $c_0=0.5$ (approximately $10 \%$ from the corresponding initial value), as shown in the bottom panel of Figure \ref{newfigure14}. The higher peak height of the kinetic energy and larger decay of the magnetic energy for $c_0=0.1$ case (along with a lower magnitude of initial Lorentz force) indicate that the MRs for $c_0=0.1$ case are more energetically efficient and generate stronger flow than $c_0=0.5$ case. We note that, with an identical MFL geometry in the vicinity of the 3D nulls, reconnections at the 3D nulls are expected to be similar for both the cases. Then, the reconnections at the QSLs for $c_0=0.1$ are expected to be more energetically efficient than the ones for $c_0=0.5$. This can be attributed to the existence of the more favorable MFL geometry and flow near the QSLs (as illustrated near the central region around $(\pi, \pi, z)$) for $c_0=0.1$ than $c_0=0.5$ \citep{sanjay-2015}.

\section{Summary and Discussions} 
\label{summary} 
 Magnetohydrodynamic simulations are presented to explore the dynamics of MFLs in the presence of 3D nulls and quasi-separatrix layers. 
The suitable initial magnetic fields $\bf{B}$ are constructed by superposing an 
exponentially decaying LFFF and a constant vertical field, with $c_0$ being 
the relative amplitude of the superposing fields. For the simulations, we select $c_0=0.1$ 
and $0.5$. The corresponding MFLs resemble 
coronal loops. The magnetic skeleton of the initial field is constructed in the form of a pair of 3D nulls, with separatrix domes and a quasi-separator between them. In addition, for $c_0=0.5$, the 3D nulls are located at lower heights in comparison to $c_0=0.1$ and the separatrix surfaces are fairly independent. While, for $c_0=0.1$ case, the separatrix domes appear to meet in the central region of the computational domain --- generating the MFL geometry of a hyperbolic flux tube with larger $Q$ values (than $c_0=0.5$) in the region. The simulated evolution is initiated by the corresponding Lorentz force.

For $c_0=0.5$, under favorable forcing,
the evolution of the magnetic skeleton documents a rotational motion of the MFLs constituting the separatrix domes. Importantly, the MFLs located in the inner vicinity of the domes do not appear to co-move with the MFLs of the domes --- causing the development of CSs at the domes. In addition, the MFLs inside the domes show rising motion toward the 3D nulls and exhibit a change in their magnetic connectivities. The development of CSs and the connectivity change of the MFLs point toward the onset of the torsional fan reconnections. 
Notably, for this case, the direction of MFL movement is found to be visibly different than the direction of plasma flow, 
indicating flipping or slipping of magnetic field lines. The weaker CSs near the QSLs than the 3D nulls show much weaker reconnection there.

In response to the initial Lorentz force, the seeming rotational motion of the MFLs corresponding
to the dome-shaped separatrix surfaces is also discerned for $c_0 = 0.1$. In this case, the CSs are found to develop under the separatrix domes. The CSs originate as the initially
parallel loops of different heights situated under the domes become non-parallel. With time, the CSs seem to extend toward the domes. Moreover, the loops appear to approach the 3D nulls and eventually alter their topology --- a clear indication of MRs at the 3D nulls. In addition, reconnection occurs at the central quasi-separator.

Further, the energy curves show the generation of stronger flow along with larger dissipation of magnetic energy for $c_0=0.1$ case (having a lower magnitude of initial Lorentz force) in comparison to $c_0=0.5$ case. This reveals that the MRs for $c_0=0.1$ are energetically more efficient and leading to a stronger outflow in comparison to $c_0=0.5$. This can be ascribed to a more favorable field line geometry and flow for $c_0=0.1$ than $c_0=0.5$, resulted from the interaction of the larger separatrix domes in the central region for $c_0=0.1$. The MRs at the favorable MFL geometry are expected to be more energetically efficient and, hence, generate large flow.

Overall, the computations document the MRs at the 3D nulls as well as at the central QSLs.
 Importantly, the results indicate that the mere presence of QSLs in initial field is not sufficient to initiate energetically efficient reconnections. The nature and magnitude of the flow 
equally crucial in commencing such reconnections.
Noticeably, the presented simulations identify the rotation of the MFLs associated with the dome-shaped fan surfaces of the 3D nulls --- also observed in the solar corona. Interestingly, under similar magnetic configuration used in the simulations, a physical scenario can be envisioned in which 
the charged particles accelerated through MRs at 3D nulls located in the corona can move along the MFLs of dome-shaped fan surfaces and potentially cause the observed circular brightening in the denser lower solar atmosphere during solar flares. In addition, the spontaneous development of the CSs can be crucial to the coronal heating. On the flip side, the presented simulations can be extended with an apt physical magnetic diffusivity to estimate the reconnection rate --- based on the MFL-aligned electric field --- at the 3D nulls and the QSLs, which is kept as a future assignment.

\begin{acks}
We acknowledge the visualisation software VAPOR (www.vapor.ucar \\
.edu)
for generating relevant graphics. A.P. acknowledges partial support of NASA grant 80NSSC17K0016 and NSF award AGS-165085. The authors thank an anonymous reviewer for providing insightful comments and suggestions to enhance the academic content as well as the presentation of the paper. 
\end{acks}



\end{article} 

\end{document}